\documentclass[pre,aps,twocolumn,showpacs,floatfix]{revtex4}
\usepackage{amsmath,amssymb,graphicx,cases}
\usepackage{epsfig}
\usepackage{textcomp}
\usepackage{epstopdf}
\usepackage{float}

\begin{document}
\title{Survival of interacting diffusing particles inside a domain with absorbing boundary}
\author{Tal Agranov}
\affiliation{Racah Institute of Physics, Hebrew University of
Jerusalem, Jerusalem 91904, Israel}
\author{Baruch Meerson}
\affiliation{Racah Institute of Physics, Hebrew University of
Jerusalem, Jerusalem 91904, Israel}
\author{Arkady Vilenkin}
\affiliation{Racah Institute of Physics, Hebrew University of
Jerusalem, Jerusalem 91904, Israel}


\begin{abstract}
Suppose that a $d$-dimensional domain
is filled with a gas of (in general, interacting)
diffusive particles with density $n_0$. A particle is absorbed whenever it reaches the
domain boundary.
Employing macroscopic fluctuation theory, we evaluate the probability ${\mathcal P}$ that no particles are absorbed during a long time $T$.
We argue that the most likely gas density profile, conditional on this event, is stationary throughout most of the time $T$. As a result, ${\mathcal P}$ decays exponentially with $T$ for a whole class of interacting diffusive gases in any dimension. For $d=1$ the stationary gas density profile and ${\mathcal P}$ can be found analytically. In higher dimensions we focus on the simple symmetric exclusion process (SSEP) and show that
$-\ln {\mathcal P}\simeq D_0TL^{d-2} \,s(n_0)$, where $D_0$ is the gas diffusivity, and $L$ is the linear size of the system.
We calculate the rescaled action $s(n_0)$  for $d=1$, for rectangular domains in $d=2$, and for spherical domains.  Near close packing of the SSEP $s(n_0)$ can be found analytically for domains of any shape and in any dimension.

\end{abstract}
\maketitle

\section{Introduction}

Diffusive lattice gases serve as useful simplified models of many stochastic spatio-temporal systems
in nature. Among them are diffusion-controlled chemical reactions: reactions which occur
quickly once the diffusing reagent particles ``find" each other in space. A simple but amazingly
rich model of this process, due to Smoluchowski \cite{Smoluchowski}, treats one
of the two reacting species as an immobile large-size minority.  The Smoluchowsi's model allows one to calculate
the expected reaction rate (that is, the expected rate of absorption of a random walker by a target).  Statistics of \emph{fluctuations} of this rate have been the subject of
numerous studies  \cite{R85,OTB89,bAH,Rednerbook,Paulbook}. When the majority molecules are treated
as noninteracting random walkers, the calculation of the effective reaction rate  and its fluctuation statistics
boils down to calculating a single-particle probability. Recently some progress has been also made in the situation when the diffusing particles interact with each other \cite{MVK,M15}.

Here we extend this line of work by
putting the walkers inside a domain which boundary is the ``target". Suppose that a gas of diffusing and, in general, interacting particles with density $n_0$ fills a $d$-dimensional
domain $\Omega$. Each particle is absorbed immediately whenever it reaches the
domain boundary $\partial \Omega$.
This simple setting is a caricature of a host of processes inside the living cell,
where a molecule needs to reach the cell membrane \cite{Bresloff}.
We assume that, on macroscopic length and time scales, the average gas density inside the
domain, $n(\mathbf{x},t)$, evolves according to a diffusion equation,
\begin{equation}
\label{difeq}
     \partial_t n = \nabla \cdot [D(n) \nabla n]
\end{equation}
with diffusivity $D(n)$ that may depend on $n$.  The boundary condition at the absorbing
domain boundary is
\begin{equation}
\label{bq}
     n(\mathbf{x}\in\partial \Omega,t) = 0.
\end{equation}
Solving Eqs.~(\ref{difeq}) and (\ref{bq}) for a given initial condition, such as  $n(\mathbf{x},t=0)=n_0$, and calculating the diffusion flux into the domain boundary, one can find  the \emph{expected} number $\bar{N}(T)$ of absorbed particles during time $T$.  In individual realizations of the underlying microscopic stochastic process, the number of absorbed particles fluctuates around  $\bar{N}(T)$, and it is interesting to determine the fluctuation statistics. In this work we will deal with an extreme limit of this statistics, corresponding  to the \emph{survival} probability  ${\mathcal P} (T)$: the probability  that not a single particle hit the
domain boundary by time $T$ which is long compared to the characteristic diffusion time
through the domain. For non-interacting diffusing particles (we will call them Random Walkers, or RWs), $D(n)=D_0=\text{const}$. In this case one obtains \cite{Paulbook}
\begin{equation}\label{RWgeneral}
-\ln {\mathcal P}_{\text{RW}}  \simeq  n_0  D_0 TV \mu_1^2,
\end{equation}
where $V$ is the domain volume, and $\mu_1$ is the lowest positive eigenvalue of the eigenvalue problem
$\nabla^2 u+\mu^2 u=0$ for the Laplace's operator inside the domain with the boundary condition $u(\mathbf{x}\in\partial \Omega) = 0$.
For a $d$-dimensional sphere of radius $R$, Eq.~(\ref{RWgeneral}) yields the well-known results
\begin{equation}\label{RWresult}
-\ln {\mathcal P}_{\text{RW}} \simeq n_0 D_0T R^{d-2} f_d ,
\end{equation}
where
\begin{numcases}
{\!\!f_d \simeq}
\frac{ \pi^2}{2}, & $d=1$, \label{survivaldecay1}\\
\pi z_1^2, & $d=2$, \label{survivaldecay2}\\
\frac{4\pi^3}{3},& $d=3$, \label{survivaldecay3}
\end{numcases}
and $z_1=2.4048\dots$ is the first positive root of the Bessel function $J_0(z)$.

The exponential decay of ${\mathcal P}_{\text{RW}}$ with time $T$, as described by Eqs.~(\ref{RWgeneral}) and (\ref{RWresult}), reflects the fact that, at long times, the \emph{single-particle} survival probability decays exponentially with time. Indeed, for a single RW,  the survival probability distribution $\rho_1(\mathbf{x},t)$ inside the domain obeys the diffusion equation $\partial_t \rho_1 = D_0 \nabla^2 \rho_1$ with the absorbing boundary condition
$\rho_1(\mathbf{x} \in \partial \Omega,t) = 0$ and a delta-function initial condition  \cite{Rednerbook}. Finding $\rho_1(\mathbf{x},t)$ and integrating it over the
domain, one obtains the single-particle survival probability as a function of time. Its long-time asymptotic describes an exponential decay with the decay rate corresponding to the lowest positive eigenvalue $\mu_1$ of the Laplace's operator. The \emph{gas} survival probability
${\mathcal P}_{\text{RW}}$ is given by the product of the survival probabilities of all independent particles inside the domain. What is left to arrive at Eq.~(\ref{RWgeneral}) is to go over to the continuum limit
by replacing a discrete sum in the exponent of ${\mathcal P}_{\text{RW}}$ by an integral. For completeness, this procedure is presented in Appendix \ref{mic}.  As a result, the exponential long-time decay of the survival probability  holds, for the independent RWs, in all spatial dimensions, as evidenced by Eq.~(\ref{RWgeneral}).

An important finding that we report here is that the exponential-in-time decay  of the survival probability $\mathcal P$ holds when the diffusing particles interact with each other, and the single-particle picture breaks down. This  non-trivial result is a consequence of the fact that the optimal gas density history, conditional on the long-time survival of all particles, is almost stationary, in any spatial dimension. We show it here by employing the Macroscopic Fluctuation Theory (MFT) \cite{MFTreview}.   For $d=1$ the stationary MFT problem is soluble for a whole family of interacting gases. In higher dimensions the solution is in general unavailable. Here we focus on a gas of particles interacting via exclusion, so as to describe e.g. diffusion-controlled chemical reactions in a crowded environment of a living cell \cite{crowded}. Specifically, we will study the symmetric simple exclusion process (SSEP). In the SSEP each particle can hop to a neighboring lattice site if that site is vacant. If it is occupied, the move is not allowed \cite{Spohn}. For the SSEP we obtain $-\ln {\mathcal P}\simeq D_0TL^{d-2} \,s(n_0)$, where $L$ is the linear size of the domain \cite{dimensions}. We calculate
the rescaled action $s(n_0)$ for several domain shapes and in different dimensions. As we show, $s(n_0)$ increases with the density $n_0$ faster than linearly, see Fig.~\ref{fff} for $d=1$, Fig.~\ref{s2d} for a rectangle in $d=2$, and Fig.~\ref{s3d} for a sphere in $d=3$. In the MFT formalism  $s(n_0)$ diverges as $n_0$ approaches unity, but
this divergence is cured when $n_0$ is sufficiently close to unity where the MFT breaks down.

The interior survival problem, considered here, has an exterior analog that is known by the name ``survival of the target". In the exterior problem
a gas of particles surrounds an absorbing domain from outside. As in the interior problem, one is interested in the probability that no particle hits the domain during time $T$. For the RWs the exterior problem was extensively studied in the past  \cite{ZKB83,T83,RK84,BZK84,BKZ86,BO87,Oshanin,BB03,BMS13}. Recently, the theory has been extended to interacting diffusive gases: for the survival probability \cite{MVK} and in the more general context of full absorption statistics \cite{M15}. As our present work, Refs. \cite{MVK,M15} employed the MFT formalism. In contrast to the interior problem, the optimal gas density history in the exterior case becomes almost stationary only for $d>2$. Furthermore,  there is a subtle but important difference
in the stationary MFT formulations of the interior and exterior problems, as we explain in the following.

In the next Section we formulate the MFT in the context of the interior survival problem.
In Sec. \ref{RWs} we apply it to the non-interacting RWs in one dimension, where ${\mathcal P}$
is known, see Eqs.~(\ref{RWresult}) and (\ref{survivaldecay1}). In this case we can solve the full time-dependent problem exactly. The solution shows that the optimal density profile for this ${\mathcal P}$ is time-independent for most of the time. The full time-dependent solution will help us identify the correct stationary formulation of the MFT problem.  In Sec. \ref{generalstat} we apply this stationarity ansatz to an arbitrary interacting diffusive gas in any dimension. This yields a stationary equation which can be simplified further upon a transformation of variable. In Sec. \ref{SSEP} we apply this procedure to the SSEP. In subsection \ref{SSEP1d} we present exact results for the SSEP in $d=1$: for the stationary optimal density profile and for the survival probability. We verify these results, in the same Section, by solving the full time-dependent MFT problem numerically. In subsection \ref{SSEP2d} we solve the stationary MFT problem for a rectangular domain, $d=2$. In subsection \ref{SSEPspherical} we study, analytically and numerically, the SSEP survival in spherical domains. In Sec. \ref{high} we identify a universal behavior of the solution in the high-density limit of the SSEP: inside a domain of any shape and in any dimension. Finally,  in Sec. \ref{otherlattice} we present a general solution of the gas survival problem  in $d=1$ which holds for a whole family of interacting diffusive gases. Section \ref{conclusion} presents a brief discussion of our results. For completeness, in Appendix \ref{mic} we calculate the survival probability of a gas of RWs from the microscopic perspective. Appendix \ref{rwnd} extends the one-dimensional solution of the MFT equations for the  RWs, presented in Sec. \ref{RWs}, to an arbitrary dimension and arbitrary domain shape.

\section{Macroscopic fluctuation theory of particle survival}
\label{MFT}

The starting point of the macroscopic fluctuation theory (MFT) \cite{MFTreview} is fluctuational hydrodynamics: a Langevin equation for the fluctuating gas density $q(\mathbf{x},t)$:
\begin{equation}
\label{Lang}
     \partial_t q = \nabla \cdot \left[D(q) \nabla q\right] +\nabla \cdot \left[\sqrt{\sigma(q)} \,\text{\boldmath$\eta$} (\mathbf{x},t)\right],
\end{equation}
where  $\text{\boldmath$\eta$} (\mathbf{x},t)$ is a zero-average Gaussian noise, delta-correlated both in space and in time. Equation (\ref{Lang}) provides an asymptotically correct large-scale description of fluctuations in a broad family of diffusive lattice gases \cite{Spohn}. At the level of fluctuational hydrodynamics, a diffusive gas is fully characterized by the diffusivity $D(q)$ and additional coefficient, $\sigma(q)$, which comes from the shot noise and is equal to twice the gas mobility coefficient \cite{Spohn}. For example, for the non-interacting RWS one has $D(q)=D_0=\text{const}$ and $\sigma(q)=2 D_0 q$, whereas for the SSEP $D(q)=D_0=\text{const}$ and $\sigma(q)=2 D_0 q(1-q)$ \cite{Spohn,dimensions}.

The MFT equations are essentially the saddle-point equations of the path-integral formulation, corresponding to the weak-noise limit of Eq.~(\ref{Lang}) \cite{MFTreview,Tailleur,DG2009b}. The MFT theory employs the typical number of particles in the relevant region of space as a large parameter. It allows to calculate the optimal path of the system: the most probable density history conditional on a specified large deviation. If the large deviation is described in terms of a spatial integral constraint, this constraint can be accommodated via the Lagrange multiplier formalism and provides a problem-specific boundary condition in time \cite{DG2009b}.

Suppose we are interested in the probability that $N$ particles are absorbed by the domain boundary by time $T$. (We will ultimately consider the limit of $N=0$.) This defines an integral constraint
on the solution:
\begin{equation}\label{number}
\int_{\Omega} d\mathbf{x}[n_0-q(\mathbf{x},T)] = N.
\end{equation}
The same type of constraint appears in the exterior problem \cite{MVK,M15}. A similar constraint also appears
in the problem of statistics of integrated current through a lattice site during a specified time \cite{DG2009b,KM_var,MS2013,MS2014,MR}.
Referring the reader to Ref. \cite{DG2009b} for a detailed derivation, we will only present here the resulting MFT equations
and boundary conditions.

The MFT equations can be written as two coupled partial differential equations for the optimal density field $q(\mathbf{x},t)$ (the ``coordinate") and the conjugate ``momentum" density field $p(\mathbf{x},t)$:
\begin{eqnarray}
  \partial_t q &=& \nabla \cdot \left[D(q) \nabla q-\sigma(q) \nabla p\right], \label{d11} \\
  \partial_t p &=& - D(q) \nabla^2 p-\frac{1}{2} \,\sigma^{\prime}(q) (\nabla p)^2, \label{d22}
\end{eqnarray}
where the prime denotes the derivative with respect to the single argument.
Equations~\eqref{d11} and \eqref{d22} are Hamiltonian,
\begin{equation}
\partial_t q = \delta H/\delta p\,, \quad
\partial_t p = -\delta H/\delta q\,.
\end{equation}
Here
\begin{equation}
\label{Hamiltonian}
H[q(\mathbf{x},t),p(\mathbf{x},t)]= \int_{\Omega} d\mathbf{x}\,\mathcal{H}
\end{equation}
is the Hamiltonian, and
\begin{equation}
\label{Ham}
\mathcal{H}(q,p) = -D(q) \nabla q\cdot \nabla p
+\frac{1}{2}\sigma(q)\!\left(\nabla p\right)^2
\end{equation}
is the Hamiltonian density. The absorbing boundary imposes zero boundary conditions for $q$:
\begin{equation}\label{bcgenq}
q(\mathbf{x}\in{\partial\Omega},t)=0.
\end{equation}
Since the values of $q$ are fixed at the boundary,  the conjugate field $p$ must vanish there \cite{MFTreview,Tailleur,MS2011}:
\begin{equation}\label{bcgenp}
p(\mathbf{x}\in{\partial\Omega},t)=0.
\end{equation}
For the RWs and SSEP this boundary condition was derived from the microscopic models \cite{Tailleur,MS2011}.
Although a general proof of Eq.~(\ref{bcgenp}) is unavailable \cite{JLprivate}, its validity has been verified
in many examples, see Refs. \cite{MFTreview,MVK,M15,MR}.

The boundary conditions in time are the following. For the density $q$ we choose a deterministic initial condition
\begin{equation}\label{t0q}
    q(\mathbf{x}\in{\Omega},t=0)=n_0.
\end{equation}
The boundary condition in time for $p(\mathbf{x},t=T)$ follows from the integral constraint (\ref{number}),
accounted for via a Lagrange multiplier \cite{DG2009b}:
\begin{eqnarray}\label{t0p}
  p(\mathbf{x}\in{\Omega},t=T)&=&\lambda,\\\nonumber
  p(\mathbf{x}\in{\partial\Omega},t=T)&=&0, \
\end{eqnarray}
where the Lagrange multiplier $\lambda$ is ultimately set by the constraint (\ref{number}).
The zero-absorption limit $N=0$, that we are interested in here, corresponds to the limit of $\lambda\to +\infty$ \cite{MR,MVK,M15}. In this limit the particle flux to the boundary vanishes at all times $0<t<T$.

Once the MFT equations with the proper boundary conditions are solved, we can calculate the action $S$ that yields $-\ln {\mathcal P}$ up to a pre-exponential factor:
\begin{eqnarray}
\label{actionmainsection1}
  -\ln {\mathcal P} \simeq S &=& \int_0^T dt\, \int_\Omega d\mathbf{x}\left(p\partial_t q-\mathcal{H}\right) \nonumber \\
  &=&\frac{1}{2}\,\int_0^T dt \int_\Omega d\mathbf{x}\,
\sigma(q)\, (\nabla p)^2.
\end{eqnarray}

\section{MFT of Random Walkers in one dimension: time-dependent solution and stationary asymptotic}
\label{RWs}
A one-dimensional domain can be set to be an interval of length $2R$, centered at the origin.
For the RWs the MFT equations become:
\begin{eqnarray}
  \partial_t q &=& D_0\partial_{x}^2 q-2D_0\partial_x (q \partial_x p), \label{qt_rw} \\
  \partial_t p &=& -D_0 \partial_{x}^2 p- D_0(\partial_x p)^2. \label{pt_rw}
\end{eqnarray}
The boundary conditions in space are
\begin{eqnarray}
q(|x|=R,t)&=&0,\label{bcq}\\
p(|x|=R,t)&=&0.\label{bcp}
\end{eqnarray}
The boundary conditions in time are
\begin{eqnarray}
q(x,t=0)&=&n_0,\label{incond}\\
p(x,t=T)&=&\lambda\,\theta(R-|x|),\label{fincond}
\end{eqnarray}
where $\theta(\dots)$ is the Heaviside step function.

\begin{figure}
\includegraphics[width=0.38\textwidth,clip=]{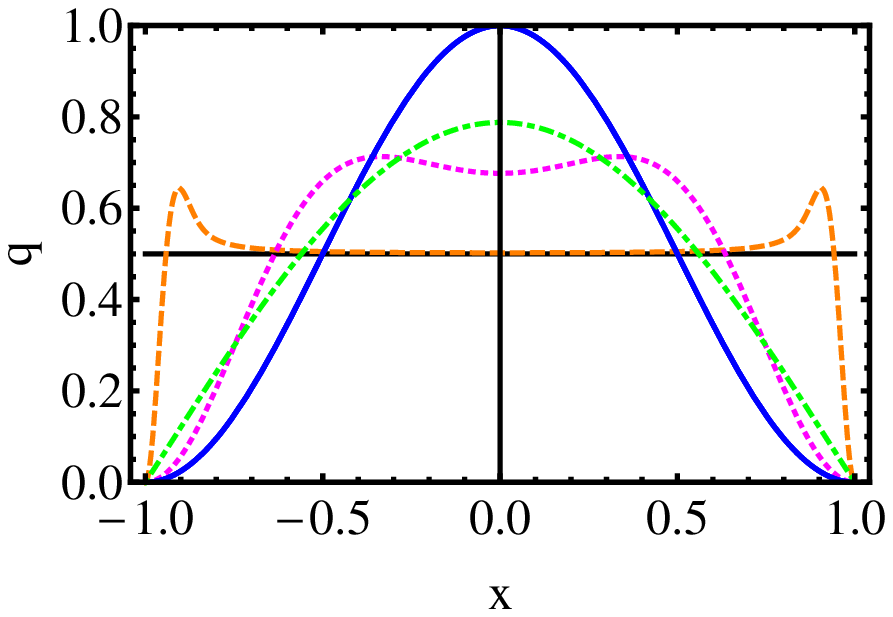}
\includegraphics[width=0.38\textwidth,clip=]{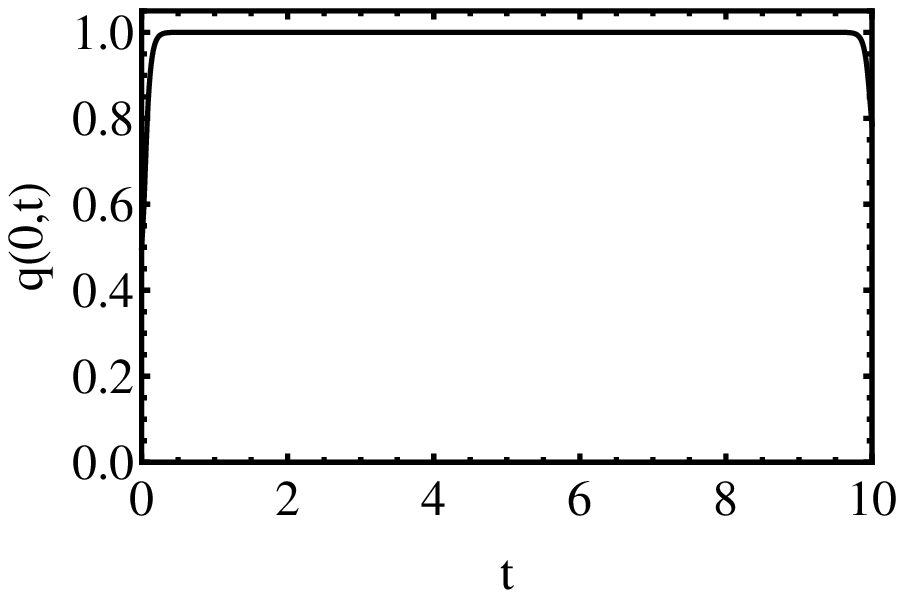}
\caption{(Color online) The exact time-dependent optimal density profile (\ref{eqqrw1}) showing the formation of a stationary solution in $d_0=1$, $T=10$ and $n_0=0.5$. Upper panel: the density profiles at times $t=0$ (solid line),  $t=0.001$ (dashed line), $t=0.05$ (dotted line), $t=1.5$, $6$ and $8.5$ (three indistinguishable solid lines), and $t=T=10$ (dash-dotted line).  The lower panel shows $q(x=0,t)$ versus time. The narrow boundary layers at $t=0$ and $t=T$ are clearly seen.}
\label{qrwprof}
\end{figure}

As many other large deviation problems for the RWs, the problem (\ref{qt_rw})-(\ref{fincond}) is exactly soluble using the Hopf-Cole transformation $Q=qe^{-p}$ and $P=e^p$, defined by the generating functional
\begin{equation}\label{ajenerating}
\int_{-R}^R dx\,F_1(q,Q)=\int_{-R}^R dx\left[q\ln(q/Q)-q\right].
\end{equation}
In the new variables the Hamiltonian density is
\begin{equation}
\label{Hamrw}
\mathcal{H}(q,p) = -D_0 \partial_x Q \partial_x P,
\end{equation}
and the action can be expressed using only the initial and final states of the system (see the Appendix of Ref. \cite{MR})
\begin{eqnarray}
\label{aaction}
-\ln {\mathcal P} \simeq S &=&\int_0^T dt\int_{-R}^R dx D_0q\left(\partial_x p\right)^2\\
&=&\int_{-R}^R dx
\left[Q\left(P\ln P -P+1\right)\right]
\big|_0^T\,.
\end{eqnarray}
\label{1d solution}
The transformed MFT equations are fully decoupled:
\begin{eqnarray}
  \partial_t Q &=& D_0\partial_{x}^2 Q, \label{aQt_rw} \\
  \partial_t P &=& -D_0 \partial_{x}^2 P, \label{aPt_rw}
\end{eqnarray}
and the transformed boundary conditions are:
\begin{eqnarray}
Q(|x|=R,t)=0,\label{abcQ}\\
P(|x|=R,t)=1,\label{abcP}
\end{eqnarray}
and
\begin{eqnarray}
Q(x,t=0)&=&\frac{n_0}{P(x,t=0)},\label{aincondrw}\\
P(x,t=T)&=&1+(e^{\lambda}-1)\theta(R-|x|).\label{afcondrw}
\end{eqnarray}
The solution of the antidiffusion equation (\ref{aPt_rw}) backward in time is obtained by integrating the
``final" condition (\ref{afcondrw}) with Green's function $G(x,x^{\prime},T-t)$, where
\begin{eqnarray}\label{green}
G(x,x^{\prime},t)&=&\frac{1}{R}\sum_{n=1}^{\infty}\sin\left[\frac{\pi n(x+R)}{2R}\right]\nonumber\\
&\times&\sin\left[\frac{\pi n(x^{\prime}+R)}{2R}\right]e^{-\frac{\pi^2 n^2 D_0 t}{4R^2}}.
\end{eqnarray}
The integration yields
\begin{eqnarray}
P(x,t)&=&1+(e^{\lambda}-1)\int_{-R}^{R} dx^{\prime} G(x,x^{\prime},T-t)\nonumber\\
&=&1+(e^{\lambda}-1)g(x,T-t), \label{aP}
\end{eqnarray}
where
\begin{equation}
\begin{split}
g&(x,t)=\int_{-R}^{R} dx^{\prime} G(x,x^{\prime},t)\\
=&\sum_{n=0}^{\infty}\frac{4}{\pi (2n+1)}\sin\left[\frac{\pi (2n+1)(x+R)}{2R}\right]e^{-\frac{\pi^2 (2n+1)^2 D_0t}{4R^2}}.
\label{g}
\end{split}
\end{equation}
Evaluating $P(x,t=0)$ from Eq.~(\ref{aP}) and using Eq.~(\ref{aincondrw}), we obtain the initial condition,
$$
Q(x,t=0)=\frac{n_0}{1+(e^{\lambda}-1)g(x,T)},
$$
for the diffusion equation~(\ref{aQt_rw}). The solution of the latter equation is
\begin{equation}
Q(x,t)=n_0 \int_{-R}^Rdx^{\prime}\frac{G(x,x^{\prime},t)}{1+(e^{\lambda}-1)g(x^{\prime},T)}. \label{aQ}
\end{equation}
Transforming back to $q$ and $p$, and taking the zero-absorption limit of $\lambda \to \infty$, we obtain:
\begin{eqnarray}
q(x,t)&=&n_0g(x,T-t)\int_{-R}^{R}dx^{\prime}\frac{G(x,x^{\prime},t)}{g(x^{\prime},T)}, \label{eqqrw1} \\
v(x,t)&=&\partial_{x}p=\partial_{x}\ln g(x,T-t), \label{eqVrw1}\\
-\ln {\mathcal P}&\simeq &-n_0\int_{-R}^{R} dx\ln g(x,T).\label{eqsrw}
\end{eqnarray}
We are interested in the long-time limit, $T\gg R^2/D_0$. A close inspection of the exact relations \eqref{green} and \eqref{g} reveals an important feature that plays a crucial role in our further analysis.  For $D_0t/R^2 \gg 1$ and $D_0(T-t)/{R^2}\gg 1$, that is, outside of narrow boundary layers (in time) of typical width $R^2/D_0$ around $t=0$ and $t=T$, the functions $G(x,x^{\prime},t)$ and $g(x,T-t)$ are well approximated by their lowest modes, $n=1$ and $n=0$, respectively:
\begin{equation}
\label{aprox1}
G(x,x^{\prime},t) \simeq  \frac{1}{R}\cos\left(\frac{\pi x}{2R}\right)\cos\left(\frac{\pi x^{\prime}}{2R}\right) e^{-\frac{\pi^2 D_0t}{4R^2}},
\end{equation}
\begin{equation}\label{aprox2}
g(x,T-t)  \simeq  \frac{4}{\pi }\cos\left(\frac{\pi x}{2R}\right) e^{-\frac{\pi^2 D_0(T-t)}{4R^2}}.
\end{equation}
Plugging these approximations into Eqs.~(\ref{eqqrw1}) and (\ref{eqVrw1}), we obtain \emph{time-independent} expressions:
\begin{eqnarray}
q(x)&= &2n_0\cos ^2\left(\frac{\pi x}{2R}\right),\label{q_rw_st}  \\
v(x)&= &-\frac{\pi}{2R}\tan\left(\frac{\pi x}{2R}\right).\label{v_rw_st}
\end{eqnarray}
That the density profile $q(x,t)$ is stationary most of the time is clearly seen in Figure~\ref{qrwprof} which shows the time-dependent solution at different times.
Notice also that it is the momentum density gradient $v(x,t)=\partial_{x}p$, and not the momentum density $p(x,t)$ itself, that stays almost stationary. This is in contrast to the exterior survival problem in $d>2$. There the density profile is also almost stationary, but it is the momentum density, and not only its gradient, that is almost stationary \cite{MVK}.

Using Eqs.~(\ref{aprox2}) and~(\ref{eqsrw}), we obtain the leading-order term of the long-time survival probability:
\begin{eqnarray}
-\ln {\mathcal P}&\simeq &n_0\frac{\pi^2 D_0T}{2R},\label{s_rw_st}
\end{eqnarray}
which is simply the action evaluated for the stationary solutions (\ref{q_rw_st}) and (\ref{v_rw_st}) on the entire interval $0<t<T$. This result agrees with Eqs.~(\ref{RWresult}) and (\ref{survivaldecay1}) as to be expected. Note that the stationary approximation remains accurate even when $D_0t/R^2$ and $D_0(T-t)/R^2$ are of order unity.
This is because the sub-leading terms in Eqs.~(\ref{eqqrw1})-(\ref{eqsrw}) include
a large factor $2\pi^2$ in the exponent. This explains the narrowness of the boundary layers on Fig.~\ref{qrwprof}, where $D_0T/R^2=10$.

Therefore, at sufficiently long times, the leading-order contribution to the survival probability of the RWs inside
a domain with an absorbing boundary
comes from stationary solutions for $q$ and $v=\partial_x p$. The stationary solution for
$v$ solves the equation
\begin{equation}
  \partial_tv=\partial_x\left(-D_0\partial_x v -D_0v^2 \right), \label{mftstatVrw}
\end{equation}
which is obtained by differentiating Eq.~(\ref{pt_rw}) in $x$.

The stationarity of the optimal density profile, which gives the leading-order contribution to the survival probability, is not unique to the one-dimensional case. We prove in Appendix \ref{rwnd} that this property holds for the RWs in any dimension and in an arbitrary domains.

Now let us revisit the problem for RWs in one dimension by directly looking for stationary solutions of Eqs.~\eqref{qt_rw} and \eqref{mftstatVrw}. Setting the mass flux in Eq.~(\ref{qt_rw}) to zero, we end up with the following set of ordinary differential equations, which we will call stationary MFT equations:
\begin{eqnarray}
  q^\prime(x)&=& 2q v, \label{eqqrw} \\
  v^{\prime}(x) &=& -v^2-\Lambda, \label{eqVrw}
\end{eqnarray}
where $\Lambda$ is a yet unknown integration constant. Note that, for the exterior problem, a similar constant vanishes \cite{MVK}.
This mathematical difference between the internal and external problems is crucial. Although its effect on the solution is immediate, its physical interpretation is somewhat elusive.

After plugging Eq.~(\ref{eqqrw}) into Eq.~(\ref{eqVrw}) we obtain a single second-order equation
\begin{equation}
 q^{\prime\prime} - \frac{(q^\prime)^2}{2q} + 2\Lambda q=0 \label{statqrw}
\end{equation}
which needs to be solved subject to the boundary condition~(\ref{bcq}) and a normalization condition, following from mass conservation:
\begin{equation}
 \frac{\int_{-R}^R dx \,q(x)}{2R}=n_0. \label{normstatqrw}
\end{equation}
To solve Eq.~(\ref{statqrw}) we make a transformation of variable
\begin{equation}
q(x)=
u^2(x) \label{transformrw}
\end{equation}
and obtain harmonic oscillator equation for $u(x)$:
\begin{equation}
u^{\prime\prime} + \Lambda u=0. \label{staturw}
\end{equation}
The solution is $u=B\sin\left[\sqrt{\Lambda}\left(x+x_0\right)\right]$, where $B$, and $x_0$ are integration constants.
The boundary conditions,
\begin{equation}
u(|x|=R,t)=0,\label{bcu}\\
\end{equation}
set $x_0=R$ and  $\Lambda=m^2\pi ^2/\left(4R^2\right)$, where $m=1,2,\dots$.
Imposing the mass conservation (\ref{normstatqrw}), we obtain a family of solutions, parameterized by $m$:
\begin{equation}
q_m(x)=2n_0\sin^2\left[\frac{m\pi(x+R)}{2R}\right].\label{eqqrwm} \\
\end{equation}
The corresponding $v$-solutions, calculated from Eq~(\ref{eqqrw}), are the following:
\begin{equation}
v_m(x)=\frac{m \pi}{2R}\cot\left[\frac{m\pi(x+R)}{2R}\right].\label{eqvrwm}
\end{equation}
Now we can calculate the action from Eq.~(\ref{actionmainsection1}):
\begin{eqnarray}
-\ln {\mathcal P_m} \simeq S_m &=& \int_0^T dt\int_{-R}^{R}dx D_0q_m v_m^2 \nonumber \\
&=&m^2n_0\frac{\pi^2 D_0T}{2R} \label{actionm}.
\end{eqnarray}
As we know from the full time-dependent solution, only the ``fundamental mode", $m=1$ is selected by the actual dynamics,
see Eqs.~(\ref{q_rw_st}) and~(\ref{v_rw_st}). Not surprisingly, this solution has the minimum action, see Eq.~(\ref{actionm}).
We argue that the same feature (selection of the lowest stationary mode) holds for the RWs in all spatial dimensions, see
Appendix \ref{rwnd}. Furthermore, it also holds for a whole class of \emph{interacting} diffusive gases. In the next section we derive stationary MFT equations for an arbitrary diffusive gas. We then solve them for the SSEP, and support our findings by a numerical solution of the full time-dependent MFT equations for the SSEP in $d=1$.

\section{Stationary MFT equations for an arbitrary diffusive gas in arbitrary dimension}
\label{generalstat}
We start with the general MFT equations (\ref{d11}) and (\ref{d22}). Taking the gradient of Eq.~(\ref{d22}), we
obtain
\begin{equation}
  \partial_t \mathbf{v} = \nabla\left[ - D(q) \nabla \cdot\mathbf{v}-\frac{1}{2} \,\sigma^{\prime}(q) \mathbf{v}^2\right], \label{Vt}
\end{equation}
where $\mathbf{v}=\nabla p$. Now we look for time-independent solutions, $q(\mathbf{x})$ and $\mathbf{v}(\mathbf{x})$ of Eqs.~(\ref{d11}) and (\ref{Vt}).  Equation~(\ref{Vt}) yields
\begin{equation}
\nabla \cdot\mathbf{v}=\frac{-\frac{1}{2} \,\sigma^{\prime}(q) \mathbf{v}^2-\Lambda}{D(q)} ,\label{Vstat}
\end{equation}
where $\Lambda$ is an integration constant to be determined later. In its turn, Eq.~(\ref{d11}) yields a zero divergence of the mass flux, so that the mass flux is a solenoidal vector field. In the survival problem, this vector field must have zero normal component to the domain boundary. Using these two properties one can show (see Appendix A of Ref. \cite{void}), that the minimum of the action is achieved when this vector field vanishes identically. Therefore, we arrive at the equation
\begin{equation}
\nabla q=\frac{\sigma(q) \mathbf{v}}{D(q)}.\label{qstat}
\end{equation}
Essentially, this equation states that, for the optimal profile,  the
fluctuation contribution to the flux exactly counterbalances the deterministic flux.
Expressing $\mathbf{v}$ from here and plugging it into Eq.~(\ref{Vstat}), we obtain a closed equation for $q$:
\begin{equation}
\nabla\cdot \left[\frac{D(q)}{\sqrt{\sigma(q)}}\nabla q\right] + \frac{\Lambda\sqrt{\sigma(q)}}{D(q)}=0,\label{qstatnd2}
\end{equation}
so that $D(q)$ and $\sigma(q)$ only enter through the combination $D/\sqrt{\sigma}$. Introduce a transformation $q=f(u)$ that satisfies the equation
\begin{equation}
\frac{D\left[f(u)\right]}{\sqrt{\sigma \left[f(u)\right]}}f^{\prime}(u)=1.\label{transform}
\end{equation}
Performing the integration (and assuming that the integral converges), we can define the function $f(u)$
implicitly:
\begin{equation}
\int_0^f dz \frac{D(z)}{\sqrt{\sigma (z)}}=u .\label{transform1}
\end{equation}
As a result of this transformation, Eq.~(\ref{qstatnd2}) becomes
\begin{equation}
\nabla^2 u +  \Lambda f^{\prime}(u)=0. \label{ustatnd}
\end{equation}
The boundary condition for $u$ is the same as for $q=f(u)$:
\begin{equation}
u(\mathbf{x}\in{\partial\Omega})=0.\label{bqu}
\end{equation}
An additional constraint on $u(\mathbf{x})$ comes from the mass conservation:
\begin{equation}\label{massgen}
  \frac{\int_{\Omega} d\mathbf{x}f(u)}{V}=n_0,
\end{equation}
where $V$ is the domain's volume.

Nonlinear equations similar to Eq.~(\ref{ustatnd}) appear in a host of physical problems. Probably the best known is the equation for stream function of an ideal incompressible fluid in two dimensions, where $-\Lambda f^{\prime}(u)$ is the vorticity \cite{Lamb,Batchelor,kaptsov}. Although the original time-dependent MFT problem has a unique solution, the stationary problem, defined by Eqs.~(\ref{ustatnd})-(\ref{massgen}), may have multiple solutions, and we will need to address the ensuing selection problem.  We already encountered this feature in the previous section, dealing with the RWs.

The stationary formulation makes it possible to deduce the scaling of the optimal density profile, and of the survival probability, with the linear system size $L$. We note that  Eq.~(\ref{ustatnd}) remains invariant upon rescaling of $\mathbf{x}$ by $L$, and of $\Lambda$ by $\sqrt{L}$. In the rescaled coordinates Eqs.~(\ref{bqu}) and (\ref{massgen}) become
\begin{eqnarray}
u(\mathbf{x}\in{\partial\tilde{\Omega}})=0.\label{bqunorm}\\
  \frac{\int_{\tilde{\Omega}} d\mathbf{x}f(u)}{\tilde{V}}=n_0,\label{massnorm}
\end{eqnarray}
where $\tilde{\Omega}$ is the rescaled domain, and $\tilde{V}$ is its volume. Extracting $\mathbf{v}$ from Eq.~(\ref{qstat}), substituting it in Eq.~(\ref{actionmainsection1}), and using Eq.~(\ref{transform}),  we obtain a simple expression for the action in terms of the new variable $u$:
\begin{eqnarray}
\label{statactionmainsection1}
  -\ln {\mathcal P} \simeq S =\frac{TL^{d-2}}{2}\int_{\tilde{\Omega}} d\mathbf{x}\left[\nabla u(\mathbf{x})\right]^2,
\end{eqnarray}
where $u$ is the solution of the rescaled problem. As we can see, the $L^{d-2}$  scaling, previously observed
for the RWs, see Eq.~(\ref{RWresult}), holds for a whole class of interacting gases. In particular,
$\ln \mathcal{P}$ is independent of $L$ for $d=2$.

We can also see how the survival probability depends on the diffusivity and mobility of the gas.  Suppose that we can express $D(q)$ and $\sigma (q)$ as $D(q)=D_0\tilde{D}(q)$ and $\sigma(q)=D_0\tilde{\sigma}(q)$, where $D_0=D(n_0)$, and $\tilde{D}, \tilde{\sigma}$ are dimensionless functions of the dimensionless density $q$. Then, from Eq.~(\ref{transform1}) we have $u=\sqrt{D_0}\,F(q)$, where $F$ is a dimensionless function of $q$ determined solely by $\tilde{D}$ and $\tilde{\sigma}$. Using this relation in Eq. (\ref{statactionmainsection1}), we obtain
\begin{equation}
  -\ln {\mathcal P} \simeq S = D_0TL^{d-2}s(n_0), \label{thesame}
\end{equation}
where $s(n_0)$ is a dimensionless function determined by the domain shape, and specialized for each model only via $\tilde{D}$ and $\tilde{\sigma}$. Comparing Eq.~(\ref{thesame}) with Eq.~(\ref{RWresult}), we see that the particle interaction
manifests itself only in the rescaled action $s(n_0)$. The same feature has been observed for the exterior survival problem \cite{MVK}.

For the RWs Eq.~(\ref{transform1}) yields $f=u^2/(2D_0)$, while Eq.~(\ref{ustatnd}) becomes the Helmholtz equation
\begin{equation}\label{Helmholtz}
\nabla^2 u+(\Lambda/D_0) \,u=0 ,
\end{equation}
which admits analytical solutions for domains of simple shapes. In one dimension this equation coincides, up to a redefinition of $\Lambda$,  with Eq.~(\ref{staturw}) of the previous section.

Let us return to Eq.~(\ref{ustatnd}). For interacting diffusive gases the function $f(u)$ is nonlinear. Still, Eq.~(\ref{ustatnd}) can be solved analytically in one dimension, and we will
exploit this fact in Sec. \ref{otherlattice}. In higher dimensions such a general solution is unavailable.  Quite a few particular solutions, in different geometries, have been found for special choices of nonlinear $f(u)$ \cite{Batchelor,Stuart,Shercliff,kaptsov,Alfimov}. Among them there is the case of $f(u)\sim \sin u$, when Eq.~(\ref{ustatnd})  becomes
a stationary sine-Gordon equation. Fortunately, this particular case describes the well-known simple symmetric exclusion process (SSEP).
As many other lattice gases, the SSEP behaves in its dilute limit
as RWs, so that $f(u\to 0)\sim u$.  Therefore, we will demand that the nonlinear solution for the SSEP cross over at low densities to the (fundamental mode) of the Helmholtz equation~(\ref{Helmholtz}).

\section{SSEP: a Stationary sine-Gordon Equation}
\label{SSEP}

Substituting $D(q)=2D_0 q$ and  $\sigma(q)=2D_0 q (1-q)$ into Eqs.~(\ref{transform1}) and (\ref{ustatnd}), we arrive at the stationary sine-Gordon equation
\begin{equation}
\nabla^2 U +  C\sin\,U=0 , \label{ustatssepnd1}
\end{equation}
where $U= \sqrt{2/D_0}\, u$,  $C= \Lambda/D_0$, and
\begin{equation}
q=f(u)=\sin^2\left(\frac{U}{2}\right).\label{transssep1}
\end{equation}
The survival probability is given by
\begin{eqnarray}\label{statactionmainsection2}
-\ln {\mathcal P} \simeq S &=&\frac{TL^{d-2}}{2}\int_{\tilde{\Omega}} d\mathbf{x}\left[\nabla u(\mathbf{x})\right]^2
\nonumber\\
&=& D_0TL^{d-2}s(n_0).
\end{eqnarray}
where
\begin{equation}\label{ssinG}
    s(n_0)= \frac{1}{4}\int_{\tilde{\Omega}} d\mathbf{x}\left[\nabla U(\mathbf{x})\right]^2 .
\end{equation}

We will now solve Eq.~(\ref{ustatssepnd1}) in several geometries.

\subsection{SSEP survival on an interval}
\label{SSEP1d}
\begin{figure}
\includegraphics[width=0.38\textwidth,clip=]{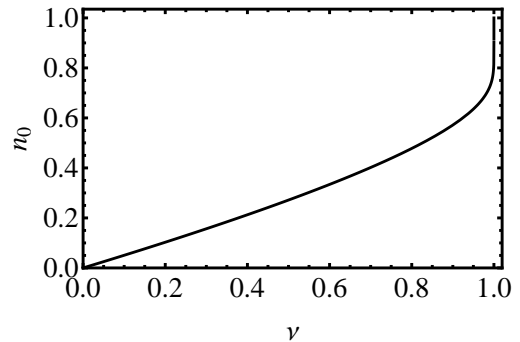}
\caption{A plot of $n_0(\nu)$ from Eq.~(\ref{n0}). }
\label{n0(nu)}
\end{figure}

As we show in section~\ref{otherlattice}, the one-dimensional case is exactly soluble in quadratures for any diffusive gas for which a stationary solution exists. In this section we find the explicit solution for the SSEP. For $d=1$ Eq.~(\ref {ustatssepnd1}) coincides with the equation of mathematical pendulum:
\begin{equation}
\frac{d ^2U}{d x^2} +  C\sin\,U=0 . \label{ustatssep1d}
\end{equation}
As for the RWs in section \ref{RWs}, we set our domain to be a segment of length $2R$ centered about the origin.
With the coordinate rescaling, presented at the end of section \ref{generalstat}, we can set $L=R=1$, whereas
Eqs.~(\ref{bqunorm}) and (\ref{massnorm}) become
\begin{eqnarray}
&&U(|x|=1)=0 ,\label{bcqp1d}\\
&& \int_{-1}^1dx\,\sin ^2\left(\frac{U}{2}\right)=2n_0 .\label{mass1d} \\
\end{eqnarray}
The general solution of Eq.~(\ref{ustatssep1d}) can be written as
\begin{equation}
U(x)=2\arcsin\left\{\sqrt{\nu} \,\text{sn}\left[\sqrt{C}(x+x_0),\nu\right]\right\}, \label{u1d}
\end{equation}
where $\text{sn}(\dots)$ is the Jacobi elliptic sine function, see e.g. Ref. \cite{elliptic}, whereas $\nu$ and $x_0$ are integration constants. The boundary condition (\ref{bcqp1d}) sets $x_0=1$ and $C=m^2\text{K}^2(\nu)$, where $\text{K}(\nu)$ is the complete elliptic integral of the first kind, and $m=1,2\dots$. The parameter $\nu$ is uniquely determined by Eq.~(\ref{mass1d}) which gives, for any $m$,
\begin{equation}\label{n0}
1-\frac{\text{E}(\nu)}{\text{K}(\nu)} =n_0,
\end{equation}
where $\text{E}(\nu)$ is the complete elliptic integral of the second kind. The plot of $n_0$ versus $\nu$ is shown in Fig. \ref{n0(nu)}. As one can see, there is a one-to-one mapping between $0<n_0<1$ and $0<\nu<1$.

\begin{figure}
\includegraphics[width=0.38\textwidth,clip=]{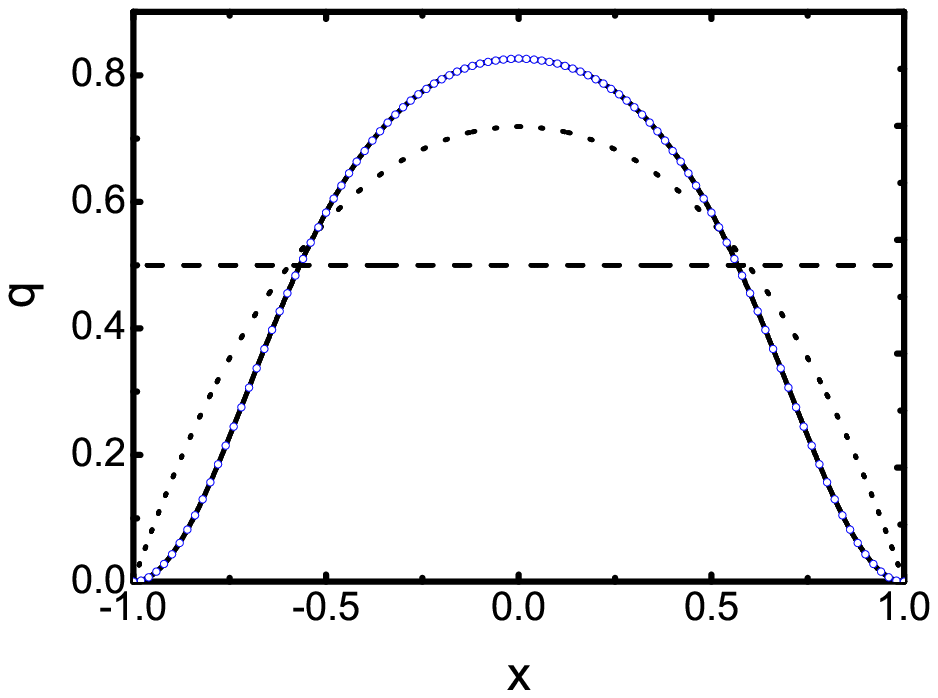}
\includegraphics[width=0.38\textwidth,clip=]{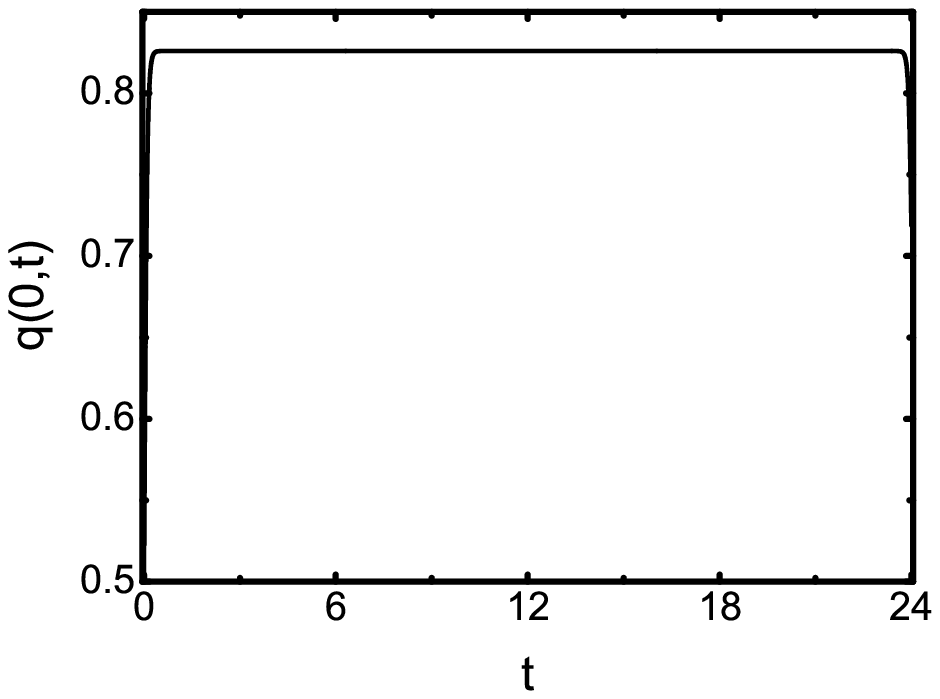}
\caption{(Color online) Formation of a stationary optimal density profile of the SSEP in one dimension, see also Fig. \ref{vx}. Upper panel: Solid line: stationary density profile~(\ref{q1}). Symbols: density profiles found numerically for $D_0=1$ and $t=6$, $12$ and $18$ (indistinguishable). Dotted line: numerical profile at $t=T=24$. Dashed line: the initial condition $q(x,0)=n_0=0.5$. Lower panel: $q(x=0,t)$ found numerically; the narrow boundary layers at $t=0$ and $t=T=24$ are clearly seen. The analytical result for $s$ from Eq.~(\ref{s}) is $s=3.5137\dots$. The numerical result from Eq.~(\ref{actionmainsection1}) is $3.545$.}
\label{3dthnum}
\end{figure}

What is left is to select the correct stationary solution out of the family of solutions parameterized by $m=1,2,\dots$.
We note that, in the dilute limit of the SSEP the solution must coincide with that for the RWs. This argument,
and the action minimization, select the fundamental mode $m=1$. Substituting $U$ in (\ref{transssep1}), we arrive at the stationary $q$-profile:
\begin{equation}\label{q1}
   q(x)=\nu \,\text{sn}^2\,\left[\text{K}(\nu) (x+1),\nu\right]=\nu \,\frac{\text{cn}^2\,\left[\text{K}(\nu) x,\nu\right]}{\text{dn}^2\,\left[\text{K}(\nu) x,\nu\right]},
\end{equation}
where $\text{cn}(\dots)$ and $\text{dn}(\dots)$ are Jacobi elliptic functions. This solution is shown by the solid line in the upper panel of Fig.~\ref{3dthnum}.

\begin{figure}
\includegraphics[width=0.38\textwidth,clip=]{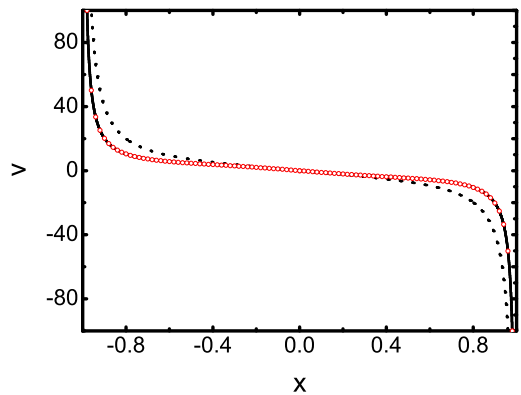}
\caption{(Color online) Formation of a stationary optimal profile of $v(x,t)$ for the SSEP in one dimension, see also Fig. \ref{3dthnum}. Solid line: the stationary profile~(\ref{V2}). Symbols: numerical profiles at $t=4$, $12$ and $18$ (indistinguishable). Dotted line: numerical profile at t=0. The parameters are $D_0=1$, $n_0=0.5$, and $T=24$.}
\label{vx}
\end{figure}

We can also calculate $v(x)=\partial_x p$ from the one-dimensional version of Eq.~(\ref{qstat}). Going back to the physical (non-rescaled) coordinate $x$, we obtain
\begin{equation}\label{V2}
v(x)=-\frac{\text{K}(\nu)\,\text{sn}\left[\frac{ \text{K}(\nu) (x)}{R},\nu\right]\text{dn}\left[\frac{ \text{K}(\nu) (x)}{R},\nu\right]}{R\,\text{cn}\left[\frac{ \text{K}(\nu) (x)}{R},\nu\right]}.
\end{equation}
Notice that
$v(x)\simeq \mp (R-|x|)^{-1}$ as $x\to \pm R$, in the same way as in the exterior survival problem \cite{MVK}. The $v(x)$-profile from Eq.~(\ref{V2}) is shown, by solid line, in Fig.~\ref{vx}.

\begin{figure}
\includegraphics[width=0.38\textwidth,clip=]{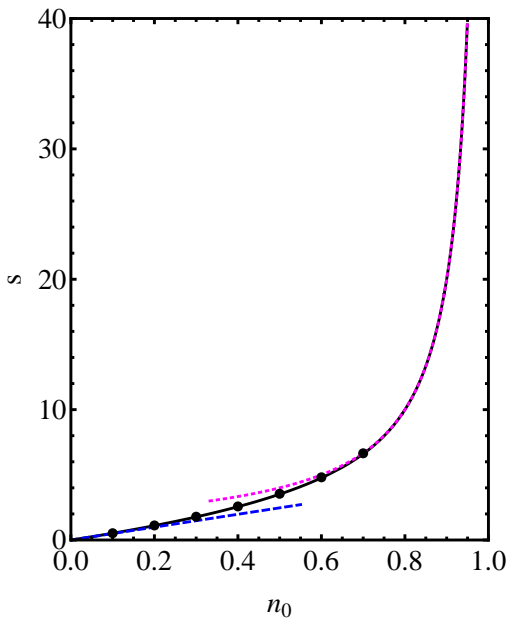}
\caption{(Color online). Solid line: the rescaled action $s(n_0)$ from Eqs.~(\ref{n0}) and (\ref{s}). Dashed line: the RW asymptotic (\ref{RW}). Dotted line: the high-density asymptotic (\ref{divergence}). Symbols: $s(n_0)$ computed from the numerical solution of the time-dependent MFT equations.}
\label{fff}
\end{figure}

Having found the stationary profile of $q$ (or $U$), we can evaluate the survival probability from Eqs.~(\ref{statactionmainsection2}) and (\ref{ssinG}):
\begin{eqnarray}
-\ln {\mathcal P}  &\simeq &   \frac{D_0 T}{4 R} \int_{-1}^{1} dx\, (U_x)^2 \nonumber\\
&=&  \frac{D_0T}{R} \int_{-1}^{1} dx\, \text{K}^2(\nu) \,\nu \,\text{cn}^2 \,\left[\text{K}(\nu) (x+1),\nu\right] \nonumber\\
&=& \frac{D_0Ts(\nu)}{R},
\label{action2}
\end{eqnarray}
where
\begin{equation}\label{s}
    s(\nu)= 2\text{K}^2(\nu) \left[\frac{\text{E}(\nu)}{\text{K}(\nu)}+\nu-1\right].
\end{equation}
Equations (\ref{n0}) and (\ref{s}) determine the rescaled action $s=s(n_0)$ in a parametric form. The low- and high-density asymptotics of $s(n_0)$ are the following:
\begin{numcases}
{\!\! s(n_0)\simeq}
\frac{\pi^2 n_0}{2}, \!\!\!\!\!\!   & $n_0\ll 1$, \label{RW}\\
\frac{2}{1-n_0},  \!\!\!\!\!\!   & $1-n_0\ll 1$, \label{divergence}
\end{numcases}
The low-density asymptotic coincides with that for the RWs, see Eqs.~(\ref{RWresult}) and (\ref{survivaldecay1}).
The high-density asymptotic formally diverges as $n_0$ approaches $1$. This divergence, however, is cured
when $n_0$ becomes very close to $1$, as explained below  in this subsection. Figure~\ref{fff} shows a plot of $s(n_0)$, alongside with the asymptotics (\ref{RW}) and (\ref{divergence}).

Also shown in Fig.~\ref{fff} are numerical results obtained by solving
the full time-dependent  MFT equations (\ref{d11}) and (\ref{d22}) for the SSEP in one dimension, using the boundary conditions in time (\ref{t0q}) and (\ref{t0p}). The numerical solution was obtained with a modified version of the iteration algorithm used in Ref.~\cite{MVK} for the exterior problem. Figures \ref{3dthnum} and \ref{vx}
show the time-dependent numerical solutions for $q$ and $v$ respectively, at different times. Apart from narrow
boundary layers at $t=0$ and $t=T$, the solutions are very close to the analytical stationary solution. The numerically evaluated rescaled action $s(n_0)$, shown in Fig. \ref{fff}, is also in very good agreement with the analytical results.

Now let us return to the high-density limit where the MFT action  (\ref{divergence}) tends to diverge. As $1-n_0\ll 1$, or $1-\nu\ll 1$, we can approximate $\text{sn}\, z =\tanh z +\mathcal{O}(1-\nu)$, and $\text{K} (\nu)\simeq (1-n_0)^{-1}$. As a result,
\begin{equation}\label{u1das}
   U(x)\simeq 2\arcsin \left[\tanh\left(\frac{1-|x|}{\delta}\right)\right],
\end{equation}
where $\delta=1-n_0\ll 1$.
The resulting density profile $q=\sin^2\left(U/2\right)$ describes two kinks, with characteristic width $\delta=1-n_0\ll 1$, located close to the ends of the interval:
\begin{equation}\label{q1das2}
q(x)\simeq
\tanh^2\left(\frac{1-|x|}{\delta} \right) .
\end{equation}
The action mostly comes from the kinks, and each kink contributes $\simeq 1/\delta$ to the action, leading to the
asymptotic (\ref{divergence}).  Now we will see how the apparent divergence of the action (\ref{divergence}) at $n_0\to 1$ is cured.
The MFT is only expected to apply when the length scales that it describes are much greater than the lattice constant
$a$. Restoring all units, we can express the kink width as $\delta \times R=(1-an_0)R$. The MFT is valid when this quantity is much greater than $a$, that is when $1-an_0\gg a/R$. On the other hand, exactly at close packing, $n_0=1/a$, the survival probability of the SSEP is
equal to the product of probabilities $P_{1,2}$ that each of the particles adjacent to the boundary does not hit the boundary during
time $T$. Each of these probabilities is $P_{1,2}=\exp(-D_0T/a^2)$, so the total survival probability is equal to $\exp(-2D_0T/a^2)$ and is of course finite.
As one can see, the crossover between the macroscopic result, $\ln \mathcal{P}_{MFT} \simeq -2D_0 T/[aR(1-an_0)]$, and the microscopic
result, $\ln \mathcal{P}=-2D_0T/a^2$ occurs at $1-an_0\sim a/R$, when the MFT theory breaks down.

As we will see in the following sections, the kink solution~(\ref{q1das2}) plays an important role in the high-density  behavior of the stationary solutions in higher dimensions, in domains of different shapes. An apparent divergence of $S$ at $n_0\to 1$ also appears there (see below), and it is also cured at the microscopic level.

\subsection{SSEP survival inside a rectangle}
\label{SSEP2d}
Here we will solve the stationary sine-Gordon equation (\ref{ustatssepnd1}) inside a rectangular domain with dimensions $L_x$ and $L_y$. After rescaling the coordinates, the dimensions of the rectangle become $1$ and $\Delta=L_y/L_x$,
in the $x$ and $y$ directions, respectively.  The equation must be solved with zero boundary conditions, whereas
Eq.~(\ref{massnorm}) reads
\begin{equation}\label{mass2d}
\frac{1}{\Delta} \int_0^{1}dx\int_0^{\Delta}dy \,q=\frac{1}{\Delta}\int_0^{1}dx\int_0^{\Delta}dy \sin ^2\left(\frac{U}{2}\right)=
n_0.
\end{equation}
As we explain shortly, an infinite family of solutions to this problem can be obtained by the method of ``generalized separation of variables" \cite{kaptsov}. In the dilute limit, one of this solution coincides with the fundamental mode of the Helmholtz's equation ~(\ref{Helmholtz}),
\begin{equation}\label{low2d}
 U=4\sqrt{n_0}\,\sin \left(\pi x\right) \sin \left(\frac{\pi y}{\Delta}\right),
\end{equation}
obeying the zero boundary conditions and the normalization condition (\ref{mass2d}), where one should replace $\sin (U/2)$ by $U/2$.  We argue, therefore, that this solution yields the true stationary optimal density profile.

\begin{figure}
\includegraphics[width=0.45\textwidth,clip=]{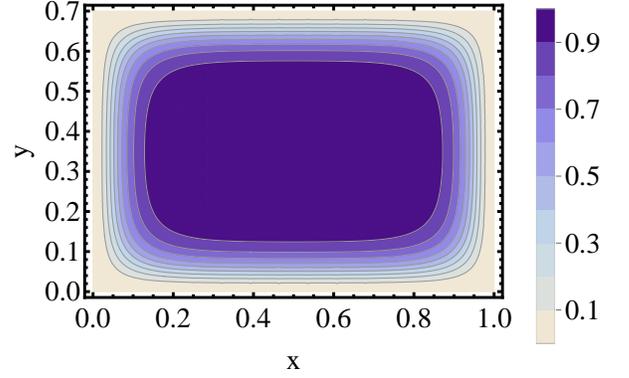}
\caption{(Color online) The stationary optimal density profile $q(x,y)$ from Eq.~(\ref{q}), corresponding to the SSEP survival in a rectangle with aspect ratio $\Delta=0.7$ and $n_0\simeq 0.67$.}
\label{recq}
\end{figure}

\begin{figure}
\includegraphics[width=0.38\textwidth,clip=]{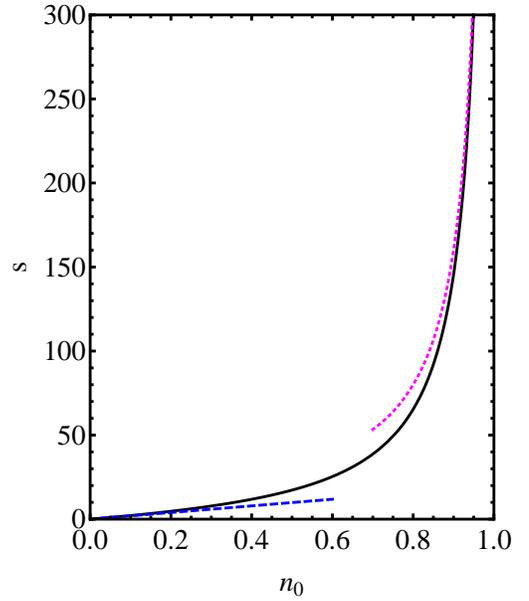}
\caption{(Color online). Solid line: the rescaled action for a square vs. $n_0$ from Eqs.~(\ref{mass2d}) and~(\ref{action2d}) with $\Delta=1$. Dashed line: the RW asymptotic (\ref{lows2d}). Dotted line: the high-density asymptotic (\ref{highs2d}).}
\label{s2d}
\end{figure}

The generalized separation of variables employs the ansatz
\begin{equation}\label{genansatz}
U(x,y)=4\arctan\left[f(x)g(y)\right]
\end{equation}
which yields two \emph{uncoupled} equations for $f$ and $g$ (see Ref. \cite{kaptsov} for a detailed derivation):
\begin{eqnarray}
  \left(f^{\prime}\right)^2&=&nf^4+mf^2+k ,\label{fxx}\\
  \left(g^{\prime}\right)^2&=&kg^4-(m+C)g^2+n,\label{gyy}
\end{eqnarray}
where  $m,n$, and $k$ are arbitrary parameters. Each of Eqs.~(\ref{fxx}) and (\ref{gyy}) describe conservation of energy
of an effective classical particle in a potential. As one can see, the particle motion is confined, and the resulting solution exhibits the correct low-density asymptotic (\ref{low2d}), if and only if $-C<m<0$, $n>0$, $k>0$, and $(m+C)^2>4kn$. In this regime of parameters the solutions for $f$ and $g$ are elliptic functions \cite{elliptic}:
\begin{eqnarray}
  f &=& \sqrt{\frac{-m\nu_1}{n(\nu_1+1)}}\,\text{sn}\left[\sqrt{\frac{-m}{\nu_1+1}}(x+c_1),\nu_1\right] ,\label{fx}\\
  g &=& \sqrt{\frac{(m+C)\nu_2}{k(\nu_2+1)}}\,\text{sn}\left[\sqrt{\frac{m+C}{\nu_2+1}}(y+c_2),\nu_2\right],\label{gy}
\end{eqnarray}
where $c_1$ and $c_2$ are the integration constants of the first order equations (\ref{fxx}) and (\ref{gyy}), and the constants $\nu_1$ and $\nu_2$ are given by $m$, $n$, and $k$ via the relations
\begin{eqnarray}
  \frac{\nu_1}{(1+\nu_1)^2} &=& \frac{kn}{m^2} ,\label{kn1}\\
  \frac{\nu_2}{(1+\nu_2)^2} &=& \frac{kn}{(m+C)^2}.\label{kn2}
\end{eqnarray}
Imposing the zero boundary condition  we obtain
\begin{eqnarray}
  \sqrt{\frac{-m}{(1+\nu_1)}} &=& 2m_1\text{K}(\nu_1), \label{bc1}\\
  \sqrt{\frac{m+C}{(1+\nu_2)}} &=& \frac{2m_2\text{K}(\nu_2)}{\Delta},\label{bc2}\\
  c_1=c_2&=&0\label{bc3},
\end{eqnarray}
where $m_1$ and $m_2$ are positive integers. Similarly to the one-dimensional case, we must put $m_1=m_2=1$.
Now we can solve Eqs.~(\ref{kn1})-(\ref{bc2}) for $\nu_1$ and obtain an expression for $fg$ in terms of $\nu_1$ alone:
\begin{equation}\label{fg}
fg=(\nu_1\nu_2)^{1/4}\,\text{sn}\left[2\text{K}(\nu_1)x,\nu_1\right]
  \text{sn}\left[\frac{2\text{K}(\nu_2)y}{\Delta},\nu_2\right] ,
\end{equation}
where $\nu_1$ and $\nu_2$ are related by the equation
\begin{equation}\label{nu122}
 \text{K}(\nu_2)^4\nu_2=\left[\text{K}(\nu_1)\Delta\right]^4\nu_1.
\end{equation}
Now we use mass conservation (\ref{mass2d}), where we substitute
\begin{equation}\label{q}
  q=\sin ^2 \left(\frac{U}{2}\right)=\frac{4(fg)^2}{\left[1+(fg)^2\right]^2}.
\end{equation}
Thus all the constants are determined implicitly.  The survival probability is given by Eq.~(\ref{statactionmainsection2}),
which we can rewrite as
\begin{eqnarray}\label{action2d}
  -\ln {\mathcal P} \simeq S &=&\frac{D_0 T}{4}\int_0^{1}dx\int_0^{\Delta}dy\,\left[\left(\partial_x U\right)^2 + \left(\partial_yU\right)^2\right]\nonumber \\
  &=&D_0 T s(n_0,\Delta).
\end{eqnarray}
As to be expected from Eq.~(\ref{thesame}), the resulting probability is independent of
the system size, but it strongly depends on the gas density. Figure~\ref{recq} shows a two-dimensional plot of $q(x,y)$
for $n_0 = 0.67$ and $\Delta=0.7$. Figure \ref{s2d}  depicts $s(n_0,\Delta=1)$ versus $n_0$, alongside with the low-
and high-density asymptotics that we will now discuss.

\subsubsection{Dilute gas, $n_0\ll 1$}

In the dilute limit, $n_0\ll 1$, our results coincide with those for the RWs. Indeed, by virtue of Eq.~(\ref{mass2d}), $U$ must be much smaller than $1$ in this limit. Therefore,   $fg$ must be also small, and we can approximate $U\simeq  4fg$, and $q\simeq 4f^2g^2$. Now, from Eq.~(\ref{mass2d}), there must be $\nu_1,\nu_2 \ll 1$. Therefore, as in the one-dimensional case, we can put $\text{sn}(\dots,\nu)\simeq \sin(\dots)$ and $\text{K}(\nu)\simeq \pi/2$ in the expression for $U$. Further, Eq.~(\ref{nu122}) yields in this limit $\nu_2\simeq \nu_1 \Delta^4$. The remaining constant $\nu_1\simeq n_0/\Delta^2$ is found from Eq.~(\ref{mass2d}). As a result, we arrive at the correct asymptotic~(\ref{low2d}). The optimal density profile in the dilute limit, back in the original coordinates, is
\begin{equation}\label{densrect}
 q\simeq 4n_0\sin^2 \left(\frac{\pi x}{L_x}\right) \sin^2 \left(\frac{\pi y}{L_y}\right).
\end{equation}
The survival probability is given by Eq.~(\ref{action2d}) with
\begin{equation}
s(n_0,\Delta) = \pi^2 n_0\left(\Delta+\frac{1}{\Delta}\right). \label{lows2d}
\end{equation}
The survival probability is maximum when the rectangle is a square.

\subsubsection{Near close packing, $1-n_0\ll 1$}

When $n_0$ is close to $1$,  the gas density is close to $1$ everywhere except in narrow boundary layers of the size $\mathcal{\delta}=1-n_0$ along the boundary. From Eq.~(\ref{q}),  $fg$ is close to $1$. As $\nu_1$ and $\nu_2$ are also close to $1$, $\text{K}(\nu_1)$ and $\text{K}(\nu_2)$ diverge as $n_0\to 1$. In this limit Eq.~(\ref{nu122}) yields $\text{K}(\nu_2)\simeq\text{K}(\nu_1)\Delta$. The explicit $n_0$-dependence can be obtained with the help of Eq.~(\ref{mass2d}): $2\text{K}(\nu_1) \simeq (1+\Delta)(\Delta\delta)^{-1}$, and $2\text{K}(\nu_2) \simeq (1+\Delta)(\delta)^{-1}$.
Using the asymptotic $\text{sn}\,z \simeq \tanh z$,
we obtain in the leading order:
\begin{eqnarray}\label{hyp}
fg&\simeq&\tanh\left[\frac{(1+\Delta)x}{\Delta\delta}\right]\tanh\left[\frac{(1+\Delta)y}{\Delta\delta}\right] \nonumber\\
&=&\frac{\tanh\left[\frac{(1+\Delta)x}{\Delta\delta}\right]+\tanh\left[\frac{(1+\Delta)y}{\Delta\delta}\right]-1}
{\tanh\left[\frac{(1+\Delta)(x+y)}{\Delta\delta}\right]} .
\end{eqnarray}
This asymptotic is valid for $x<1/2$ and $y<\Delta/2$; it can be extended to the other three quarters of the rectangle by obvious reflections.
Away from the domain corners we can replace $\tanh(\dots)$ in the denominator of Eq.~(\ref{hyp}) by unity.
Using the resulting ``approximate product rule" in Eq.~(\ref{q}), we obtain, after some algebra, the ``kink" asymptotic of $q(x,y)$ away from the domain corners:
\begin{equation}\label{qaprox}
  q\simeq \tanh^2\left[\frac{2(1+\Delta)x}{\Delta\delta}\right]\tanh^2\left[\frac{2(1+\Delta)y}{\Delta\delta}\right],
\end{equation}
where  $x<1/2$ and $y<\Delta/2$, and reflected formulas in the other three rectangle quarters.
These asymptotics describe kinks with the characteristic width $\ell=\Delta\delta/[2(1+\Delta)]$.  By analogy with one dimension, the action per unit length along the boundary is, in the leading order, $1/\ell$. Multiplying this expression by the perimeter $2(1+\Delta)$, we extract the asymptotics
\begin{equation}\label{highs2d}
  s(n_0,\Delta)\simeq \frac{4}{1-n_0} \left(\Delta+\frac{1}{\Delta}+2\right).
\end{equation}
Again, the survival probability is maximum when the rectangle is a square.

\subsubsection{Very long rectangle, $\Delta\gg 1$}

Here, sufficiently far from the edges $y=0$ and $y=\Delta$, $U(x,y)$ is almost independent of $y$, and close to the one-dimensional solution~(\ref{u1d}). Therefore, the rescaled action $s(n_0,\Delta)$ is approximately equal to
\begin{equation}\label{asdelta}
 s(n_0,\Delta\gg 1)\simeq 2\Delta s_{1d}(n_0),
\end{equation}
where $s_{1\text{d}}(n_0)$ is described by Eqs.~(\ref{n0}) and (\ref{s}). The factor $\Delta$ is due
to additional integration along $y$, and the factor $2$ appears because the one-dimensional result~(\ref{s}) was obtained for a
segment of length $2$, not $1$.

\subsection{SSEP survival inside a sphere}
\label{SSEPspherical}
Here the stationary optimal density profile depends only on the radial coordinate $r$, and Eq.~(\ref{ustatssepnd1}) becomes
\begin{equation}
\nabla_r^2 U +  C\sin U=0 , \label{ustatssepnd1r}
\end{equation}
where $\nabla_r^2U(r)=U_{rr} + (d-1)U_r/r$ is the radial Laplacian in $d$ dimensions.
Upon the coordinate rescaling $r \to r/R$, we need to solve the stationary sine-Gordon equation~(\ref{ustatssepnd1r}) inside a sphere of unit radius. The boundary conditions, and the normalization condition, are
\begin{eqnarray}
U^{\prime}(r=0)&=&U(r=1)=0 ,\label{bcqpnd}\\
d\int_0^1drr^{d-1}q(r)&=&d\int_0^1drr^{d-1}\sin ^2\left(\frac{U}{2}\right)=n_0 .\label{massnd} \nonumber\\
\end{eqnarray}
Then, from Eqs.~(\ref{statactionmainsection2}) and (\ref{ssinG}), we obtain $-\ln {\mathcal P} \simeq S=D_0TR^{d-2}s(n_0)$,
where
\begin{equation}\label{actionmain}
   s(n_0)=\frac{\Omega_d}{4}\int_0^1drr^{d-1}(U_r)^2,
\end{equation}
and $\Omega_d$ is the surface area of the $d$-dimensional unit sphere.
In the absence of general analytic solution of Eq.~(\ref{ustatssepnd1r}) for $d>1$, one can solve this equation numerically, and also explore analytically the low- and high-density limits.

The first-order term of the density expansion of $s(n_0)$ corresponds to the RWs, see Appendix \ref{rwnd}. The next, $n_0^2$-term  can be obtained by treating the $q^2$  term of the MFT Hamiltonian of the SSEP perturbatively, similarly to how it was done in the exterior problem \cite{MVK}. For example, for $d=3$ the resulting correction is
\begin{equation}\label{correction}
    \delta s(n_0) = 4 \pi \int_0^1 dr\,r^2 q_{\text{RW}}^2(r) v_{\text{RW}}^2 (r),
\end{equation}
where $q_{\text{RW}}(r)$ and $v_{\text{RW}}(r)$ are the stationary optimal profiles for the RWs, given by Eqs.~(\ref{q_rw_st3d})
and (\ref{v_rw_st3d}) of Appendix B.  Evaluating the integral, we obtain $\delta s(n_0) = \alpha \,n_0^2$, where
$$
 \alpha = \frac{8\pi^4}{27}\left[2\,\text{Si}\,(2\pi)-\,\text{Si}\,(4 \pi)\right]=38.7945\dots ,
$$
and
$$\text{Si}\,(z)=\int_0^z \frac{\sin t}{t} \,dt$$
is the sine integral function. The resulting low-density asymptotic, for $d=3$, is
\begin{equation}\label{lowdens}
    s(n_0)\simeq \frac{4\pi^3}{3}\, n_0+\alpha \,n_0^2.
\end{equation}

Near close packing, $1-n_0\ll 1$, the gas density $q(r)$ drops from a value close to $1$ to zero in a narrow boundary layer, of width $\mathcal{O}(\delta)$ near the sphere $r=1$. Correspondingly, $U(r)=2\arcsin \sqrt{q(r)}$ rapidly drops from a value close to $\pi$ to zero. As a result, we can neglect the first-derivative term in the radial Laplacian. This brings us back to the equation  $U_{rr} + C\sin U=0$ that we considered in Sec.~\ref{SSEP1d}. The boundary conditions (\ref{bcqpnd}) are also the same as in the one-dimensional case, see Eq.~(\ref{bcqp1d}). The only difference is in the normalization condition, Eq.~(\ref{massnd}) which introduces the factor $d$. As a result,
\begin{equation}\label{qaproxnd2}
   U(r)\simeq 2\arcsin\left\{\tanh\,\left[\frac{d(1-r)}{1-n_0}\right]\right\},
\end{equation}
and
\begin{equation}\label{qaproxnd22}
   q(r)=\sin^2 \left(\frac{U}{2}\right)\simeq \tanh^2\,\left[\frac{d(1-r)}{1-n_0}\right].
\end{equation}
Using Eq.~(\ref{qaproxnd2}), we obtain the high-density asymptotic
\begin{equation}\label{actionaproxnd}
  s(n_0) =\frac{\Omega_d}{4}\int_0^1drr^{d-1}(U_r)^2\simeq \frac{d \Omega_d}{1-n_0},
\end{equation}
In particular, for $d=3$,
\begin{equation}\label{divergence3d}
s_{3d}(n_0)\simeq \frac{12\pi}{1-n_0},  \quad 1-n_0\ll 1.
\end{equation}

\begin{figure}
\includegraphics[width=0.38\textwidth,clip=]{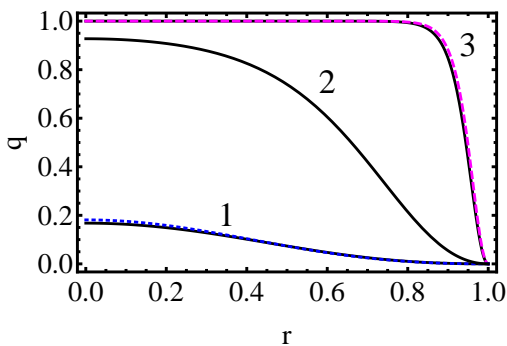}
\caption{(Color online). Stationary optimal density profiles for the SSEP survival inside a sphere, $d=3$. Solid lines: profiles, obtained by numerically solving Eq.~(\ref{ustatssepnd1r}) with three different values of $C$, corresponding to three different values of $n_0$, and using Eq.~(\ref{transssep1}). Curve 1: $C=10.4$ and $n_0\simeq 0.027$. Dotted line: the asymptotic profile~(\ref{q_rw_st3d}). Curve 2: $C=20$ and $n_0\simeq 0.31$. Curve 3: $C=300$ and $n_0\simeq 0.82$. Dashed line: the asymptotic profile~(\ref{qaproxnd2}).}
\label{q3d}
\end{figure}
\begin{figure}
\includegraphics[width=0.35\textwidth,clip=]{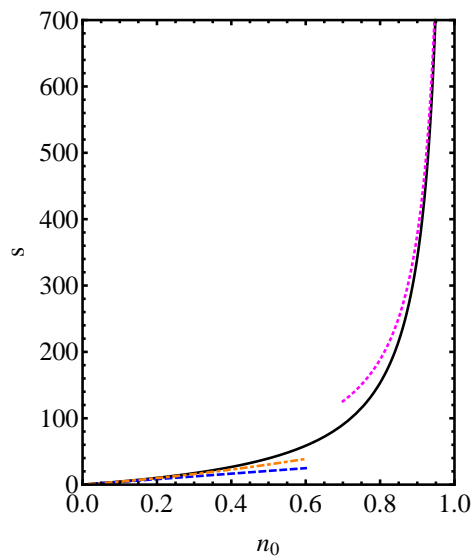}
\caption{(Color online). Solid line: the rescaled action $s(n_0)$ for a sphere, $d=3$, vs. $n_0$, obtained by numerically solving Eq.~(\ref{ustatssepnd1r}) and using Eq.~(\ref{actionmain}). Also shown are the low-density asymptotic~(\ref{lowdens})
(dash-dotted line) and its linear term  (dashed line).  Dotted line: the high-density asymptotics~(\ref{divergence3d}).}
\label{s3d}
\end{figure}

For an arbitrary density $n_0$  Eq.~(\ref{ustatssepnd1r}) can be solved numerically: either by the shooting method or by artificial relaxation.  Examples of such solutions for $d=3$ are shown in Fig.~\ref{q3d}. Figure~\ref{s3d} shows the numerically found $s_{3d}(n_0)$,  alongside with the asymptotic (\ref{lowdens}), its linear term only,  and asymptotic~(\ref{divergence3d}).

\subsection{SSEP survival in arbitrary domains near close packing}
\label{high}

Consider a domain of arbitrary shape, in any dimension. As $n_0$ approaches $1$, the stationary optimal density field $q$ stays very close to $1$ across most of the domain, and drops to $0$ in a narrow boundary layer of characteristic width $\delta =1-n_0$ along the domain boundary. The density derivatives in the directions parallel to the boundary are negligibly small compared to the density derivative across the boundary. Therefore,  we can approximate the Laplacian in Eq.~(\ref{ustatssepnd1}) by a one-dimensional one,  $\partial^2U/\partial r_{\perp}^2$, where $r_{\perp}$ is a local coordinate normal to the domain boundary.
As a result, the problem becomes effectively one-dimensional, and the solution of Eq.~(\ref{ustatssepnd1})  is a one-dimensional kink, $q(\mathbf{x})\simeq \tanh^2\left(\sqrt{C}\,r_{\perp} \right)$ (we set $r_{\perp}$ to vanish at the boundary). The action, Eq.~(\ref{statactionmainsection2}), mostly comes from the boundary layer. In the leading order, the action per unit surface across the boundary is equal to $\sqrt{C}$. The total action is therefore $\tilde{A}\sqrt{C}$, where $\tilde{A}$ is the normalized domain's surface area.  $\sqrt{C}$ is determined by  the mass conservation: $\sqrt{C}\simeq \tilde{A}/(\tilde{V}\delta)$, where $\tilde{V}$ is the normalized domain's volume. The final result is
\begin{equation}\label{highs}
  -\ln {\mathcal P} \simeq S \simeq D_0TL^{d-2}\frac{\tilde{A}^2}{\tilde{V}\left(1-n_0\right)}=\frac{A^2 D_0T}{V\left(1-n_0\right)}.
\end{equation}
This expression is in agreement with our high-density results (\ref{divergence}), (\ref{highs2d}), and (\ref{actionaproxnd}).

\section{Survival of a general diffusive gas on an interval}
\label{otherlattice}
For a general diffusive gas in one dimension the stationary density profile and the survival probability can be found in quadrature.  Indeed, for arbitrary $D(q)$ and  $\sigma(q)$ in one dimension,  Eq.~(\ref{ustatnd}) reads
\begin{equation}
u^{\prime\prime}+  \Lambda f^{\prime}(u)=0 , \label{ustat1d}
\end{equation}
where $f(u)$ is defined by Eq.~(\ref{transform1}). By virtue of the scaling properties of the problem, it suffices to solve Eq.~(\ref{ustat1d}) on the interval $|x|<1$,  with the same additional conditions for $u$ as stated in Eqs.~(\ref{bcqp1d}) and (\ref{mass1d}) for $U$. Equation~(\ref{ustat1d}) describes the motion of an effective classical particle with unit mass ($u$ is the ``particle coordinate", $x$ is ``time") in the potential $V(u)=\Lambda f(u)$. Let us denote the particle energy by $\Lambda\nu$. Energy conservation yields a first-order equation:
\begin{equation}\label{Qprimegen}
u^{\prime}=\pm\sqrt{2 \Lambda \left[\nu-f(u)\right]}
\end{equation}
Solving it with the boundary condition $u(x=-1)=0$, we obtain:
\begin{equation}\label{ueq}
\int_{0}^{u}\frac{dh}{\sqrt{\nu-f(h)}}=\sqrt{2\Lambda}\,(x+1).
\end{equation}
Changing the integration variable to $z=f(h)$ and using Eq.~(\ref{transform}), we obtain
\begin{equation}\label{qeq}
\int_{0}^{q}dz\,\frac{D(z)}{\sqrt{(\nu-z)\sigma(z)}}=\sqrt{2\Lambda}\,(x+1).
\end{equation}
Now we demand that $q=\nu$ at $x=0$, and express $\Lambda$ through $\nu$:
\begin{equation}\label{A1}
  \Lambda=\frac{1}{2}\left[\int_0^{\nu}\frac{dz\,D(z)}{\sqrt{(\nu-z) \,\sigma(z)}}\right]^2.
\end{equation}
An additional condition comes from mass conservation:
\begin{eqnarray}
\label{conservation}
  2n_{0} &=& \int_{-1}^{1}q(x)dx=2\int_{0}^{\nu}q\,\frac{dx}{dq}\left(q\right)\,dq \nonumber \\
  &=& 2\int_{0}^{\nu}dq\frac{q \,D(q)}{\sqrt{2\Lambda(\nu-q)\sigma(q)}}.
\end{eqnarray}
Equations~(\ref{A1}) and (\ref{conservation}) yield the dependence of $n_0$ on $\nu$:
\begin{equation}\label{n0gen}
n_0(\nu)=\frac{\int_{0}^{\nu}dq\frac{q D(q)}{\sqrt{(\nu-q)\sigma(q)}}}{\int_{0}^{\nu}dq\frac{D(q)}{\sqrt{(\nu-q)\sigma(q)}}}.
\end{equation}
Now we can evaluate the survival probability:
\begin{equation}\label{actn1}
-\ln {\mathcal P} \simeq \frac{T}{2R}\int_{-1}^{1}dx \, \left(\frac{d u}{d x}\right)^2.
\end{equation}
By virtue of Eq.~(\ref{Qprimegen}) and the definition $q=f(u)$, we obtain
\begin{equation}\label{actn2}
-\ln {\mathcal P} \simeq \frac {\Lambda T}{R}\int_{-1}^{1} [\nu-q(x)] dx =\frac{2\Lambda T(\nu-n_{0})}{R},
\end{equation}
where we have again used  $\int_{-1}^{1} q(x)dx = 2n_{0}$.
Using Eq.~(\ref{A1}), we can rewrite Eq.~(\ref{actn2}) as
\begin{equation}\label{actn3}
-\ln {\mathcal P} \simeq \frac{T}{R}\,s(n_0),
\end{equation}
where the rescaled action $s=s(n_0)$ is given in a parametric form by the equation
\begin{equation}\label{LDFgen}
s(\nu)= \left[\int_0^{\nu}\frac{dq\,D(q)}{\sqrt{(\nu-q) \,\sigma(q)}}\right]^2 \left[\nu-n_0(\nu)\right]
\end{equation}
and Eq.~(\ref{n0gen}). When specialized to the SSEP, Eqs.~(\ref{qeq}) and (\ref{LDFgen}) yields (\ref{q1}) and (\ref{s}) respectively.

With these general results at hand, we can investigate the survival properties of a whole class of diffusive gases
with known $D(q)$ and $\sigma(q)$ in one dimension: on the condition that the integral in the denominator of Eq.~(\ref{n0gen}) converges.

\section{Conclusions and Discussion}
\label{conclusion}

We dealt in this work with the survival of a gas of interacting diffusive particles inside a domain with absorbing boundary. Employing the MFT formalism, we evaluated the long-time survival probability of the gas and its optimal density history conditional on the survival. We found that this optimal density history is stationary during most of the process. As a consequence, the survival probability decays exponentially in time:  inside domains of any shape in all dimensions. As we showed, the solution of the long-time survival problem reduces to solving
a nonlinear Poisson equation, where the nonlinear term is determined by $D(q)$ and $\sigma(q)$.  In one dimension, this problem is soluble exactly for a whole class of diffusive gases. For the SSEP the nonlinear Poisson equation becomes
a stationary sine-Gordon equation, and we solved it in different geometries and dimensions.

The dilute limit of the SSEP corresponds to non-interacting random walkers (RWs), where the problem reduces to finding the lowest positive eigenvalue $\mu_1$ of the Laplace's operator inside the domain \cite{Paulbook}, see Eq.~(\ref{RWgeneral}).  Near close packing the problem becomes effectively one-dimensional and can be approximately solved for any domain shape and any dimension.  What is the optimal domain shape, for a fixed number of particles and fixed volume of the domain, that maximizes the chances of long-time survival? Interestingly, both in the dilute limit of the SSEP, and near close packing, the optimal domain shape is the sphere.  Indeed, in the dilute limit the minimum value of $\mu_1$ is achieved for the sphere, as guaranteed by the Rayleigh-Faber-Krahn theorem  \cite{ball,Chavel}. In its turn, near the close packing, the sphere is the minimizer of the surface area $A$ at fixed volume, see Eq.~(\ref{highs}). A natural conjecture is that
the sphere maximizes the survival probability of the SSEP at any gas density, but we do not have a proof. For other diffusive gases we do not know the minimizing domain shape.

It would be interesting to apply our approach to the ``narrow escape problem", where there is a small hole in the \emph{reflecting} boundary of the domain. The survival probability \cite{hole}  and the mean escape time \cite{Holcman} of a single RW in this system  have been extensively studied. The MFT formalism can give an interesting insight into fluctuations in the escape of \emph{interacting} particles.

Another interesting extension would address the full absorption statistics: evaluating the probability that a specified number of particles are absorbed by time $T$. An exterior variant of this problem has been recently considered, for the SSEP, in Ref.~\cite{M15}.

Finally, we emphasize that stationarity of the optimal gas density profile is a major simplifying factor in the large-deviation problem we have considered here.  Cases of time-independence of the optimal gas density history are also encountered in other large-deviation settings in lattice gases \cite{Bodineau,Hurtado1,Hurtado2,MVK,M15}, and they are intimately related to the ``additivity principle" put forward by Bodineau and Derrida \cite{Bodineau}.

\section*{Acknowledgments}
We thank O.V. Kaptsov for sending us the book \cite{kaptsov} that he coauthored, and P.L. Krapivsky for a useful discussion of the Rayleigh-Faber-Krahn theorem.
This research was supported by grant No.\ 2012145 from the
United States--Israel Binational Science Foundation (BSF).

\appendix
\section{Survival probability of Random Walkers from a microscopic perspective}\label{mic}
For completeness, we present here a brief microscopic derivation of the survival probability of a gas of non-interacting Random Walkers
(RWs) inside a domain $\Omega$. This quantity can be obtained from a single-particle survival probability:
\begin{equation}\label{rwp}
-\ln {\mathcal P(T)}_{\text{RW}}  = - \sum_i \ln g(\mathbf{x}_i,T),
\end{equation}
where $g(\mathbf{x}_i,T)$ is the survival probability up to time $T$ of a particle initially positioned at $\mathbf{x}_i$ , and the sum is over all particles. Therefore, one needs to calculate $g(\mathbf{x}_i,T)$ and perform the summation.

\subsection{Calculating $g(\mathbf{x}_i,T)$}
The single-particle survival probability $g(\mathbf{x}_i,T)$ is given by the expression
\begin{equation}\label{sp}
  g(\mathbf{x}_i,T)=\int_{\Omega} d\mathbf{x}\,\rho_1(\mathbf{x},T,\mathbf{x}_i),
\end{equation}
where $\rho_1(\mathbf{x},t,\mathbf{x}_i)$ is the probability distribution of the particle position, given its (deterministic) initial position at $\mathbf{x}_i$. This probability distribution obeys the diffusion equation \cite{R85,OTB89,bAH,Rednerbook}:
\begin{equation}\label{dif}
\partial_t \rho_1 = D_0 \nabla^2 \rho_1
\end{equation}
with the absorbing boundary condition:
\begin{equation}\label{bcrw}
  \rho_1(\mathbf{x} \in \partial \Omega,t) = 0.
\end{equation}
The initial condition is
\begin{equation}\label{incon}
  \rho_1(\mathbf{x},t=0)=\delta^d(\mathbf{x}-\mathbf{x}_i),
\end{equation}
where $\delta^d$ is the $d$-dimensional Dirac delta-function.
The solution to Eq.~(\ref{dif}) is the Green's function of the diffusion equation:
\begin{equation}\label{agreennd}
\rho_1(\mathbf{x},t,\mathbf{x}_i)=G(\mathbf{x},\mathbf{x}_i,t)=\sum_{n=1}^{\infty}\Psi_n(\mathbf{x})\Psi_n(\mathbf{x}_i)e^{-\mu_n^2 D_0t}.
\end{equation}
Here $\Psi_n$ and $\mu_n$ are the normalized eigenfunctions and eigenvalues of the eigenvalue problem
$\nabla^2 u+\mu^2 u=0$ for the Laplace's operator inside the domain with the boundary condition $u(\mathbf{x}\in\partial \Omega) = 0$. We order the eigenvalues by their magnitude $\mu_1<\mu_2<\mu_3<\dots$.
Plugging Eq.~(\ref{agreennd}) into Eq.~(\ref{sp}) one obtains \cite{Paulbook}:
  \begin{eqnarray}
g(\mathbf{x}_i,T)&=&\int_{\Omega} d\mathbf{x}\,G(\mathbf{x},\mathbf{x}_i,t) \nonumber\\
&=& \sum_{n=1}^{\infty}\Psi_n(\mathbf{x}_i)\int_{\Omega} d\mathbf{x}\,\Psi_n(\mathbf{x})e^{-\mu_n^2 D_0T}
\label{fnd}
\end{eqnarray}

\subsection{Evaluating the sum in Eq.~(\ref{rwp})}
When the number of particles in the domain $\Omega$ is very large, the sum in (\ref{rwp}) can be approximated by the integral:
\begin{equation}
  -\ln {\mathcal P(T)}_{\text{RW}}  \simeq  -n_0\int_\Omega d\mathbf{x}^{\prime}\ln\left[g(\mathbf{x}^{\prime},T)\right],\label{aeqsrwnd}
\end{equation}
Furthermore, at times much longer than the characteristic diffusion time, the infinite sum in Eq.~(\ref{fnd}) can be approximated by its first term:
\begin{equation}\label{gaprox}
  g(\mathbf{x},T)  \simeq  \Psi_1(\mathbf{x})\int_{\Omega} d\mathbf{x^{\prime}}\Psi_1(\mathbf{x^{\prime}})e^{-\mu_1^2 D_0T}.
\end{equation}
Plugging this approximation into Eq.~(\ref{aeqsrwnd}), we obtain the long-time asymptotic of the survival probability
presented in Eq.~(\ref{RWgeneral}).

\section{Solving the MFT equations for the RWs in higher dimensions}
\label{rwnd}
The one-dimensional solution, presented in section \ref{RWs}, can be generalized to any simply-connected domain in arbitrary dimension.
Consider the MFT equations (\ref{d11}) and (\ref{d22}) for the RWs:
\begin{eqnarray}
  \partial_t q &=& \nabla \cdot \left[D_0 \nabla q-2D_0q \nabla p\right], \label{ad1} \\
  \partial_t p &=& - D_0 \nabla^2 p-D_0 (\nabla p)^2, \label{ad2}
\end{eqnarray}
The Hamiltonian density is
\begin{equation}
\label{aHam}
\mathcal{H}(q,p) = -D_0 \nabla q\cdot \nabla p
+D_0q\!\left(\nabla p\right)^2 .
\end{equation}
The absorbing boundary conditions are described by Eq.~(\ref{bcgenq}) and (\ref{bcgenp}).
The boundary conditions in time are given by Eqs.~(\ref{t0q}) and (\ref{t0p}).

As in one dimension, we solve the problem, using the Hopf-Cole transformation $Q=qe^{-p}$ and $P=e^p$, with the generating functional
\begin{equation}\label{ajeneratingnd}
\int_{\Omega} d\mathbf{x}F_1(q,Q)=\int_{\Omega} d\mathbf{x}\left[q\ln(q/Q)-q\right].
\end{equation}
In the new variables the Hamiltonian density is
\begin{equation}
\label{Hamrwnd}
\mathcal{H}(q,p) = -D_0 \nabla Q\cdot \nabla P,
\end{equation}
and, again, the action can be expressed through the initial and final states of the system:
\begin{eqnarray}
\label{aactionnd}
-\ln {\mathcal P}_{\text{RW}} &\simeq& S =\int_0^T dt\int_\Omega d\mathbf{x} D_0q\left(\nabla p\right)^2\\
&=&\int_\Omega d\mathbf{x}\left[Q\left(P\ln P -P+1\right)\right]\big|_0^T
\end{eqnarray}
The transformed MFT equations are decoupled:
\begin{eqnarray}
  \partial_t Q &=& D_0\nabla^2 Q, \label{aQt_rwnd} \\
  \partial_t P &=& -D_0 \nabla^2 P. \label{aPt_rwnd}
\end{eqnarray}
The transformed boundary conditions, in space and in time, are:
\begin{eqnarray}
&&Q(\mathbf{x}\in{\partial\Omega,t})=0, \label{abcQnd}\\
&&P(\mathbf{x}\in{\partial\Omega,t})=1, \label{abcPnd}\\
&&Q(\mathbf{x},t=0)=\frac{n_0}{P(\mathbf{x},t=0)}, \label{aincondnd}
\end{eqnarray}
and
\begin{numcases}
{P(\mathbf{x},t=T)=}
e^{\lambda}, & $\mathbf{x}\in{\Omega}$\nonumber\\
1. & $\mathbf{x}\in{\partial\Omega}$
\end{numcases}
Solving the anti-diffusion equation (\ref{aPt_rwnd}), we obtain
\begin{equation}
P(\mathbf{x},t)=1+(e^{\lambda}-1)g(\mathbf{x},T-t), \label{aPnd}
\end{equation}
where $g$ is defined in Eq.~(\ref{fnd}). Evaluating $P(\mathbf{x},t=0)$, we obtain the initial condition for the diffusion equation~(\ref{aQt_rwnd}):
$$
Q(\mathbf{x},t=0)=\frac{n_0}{1+(e^{\lambda}-1)g(\mathbf{x},T)}.
$$
The resulting solution of Eq.~(\ref{aQt_rwnd}) is
\begin{equation}
Q(\mathbf{x},t)=n_0 \int_{\Omega} d\mathbf{x^{\prime}} \frac{G(\mathbf{x},\mathbf{x}^{\prime},t)}{1+(e^{\lambda}-1)g(\mathbf{x}^{\prime},T)}. \label{aQnd}
\end{equation}
Now we calculate the action using Eq.~(\ref{aactionnd}). After some algebra, and taking the zero-absorption limit of $\lambda \to \infty$, we arrive at  Eq.~(\ref{aeqsrwnd}),
which describes the continuum approximation of the exact microscopic result (\ref{rwp}).
Transforming back to $q$ and $p$, and taking the limit of $\lambda \to \infty$, we obtain:
\begin{eqnarray}
q(\mathbf{x},t)&=& n_0 g(\mathbf{x},T-t)\int_\Omega d\mathbf{x^{\prime}} \frac{G(\mathbf{x},\mathbf{x}^{\prime},t)}{g(\mathbf{x}^{\prime},T)}, \label{aeqqrwnd}\\
\mathbf{v}(\mathbf{x},t)&=&\nabla p=\nabla \ln g(\mathbf{x},T-t).  \label{aeqVrwnd}
\end{eqnarray}
Being interested in long times, we observe that,  outside the boundary layers of width $L^2/D_0$ around $t=0$ and $t=T$, one can approximate expressions Eq.~(\ref{fnd}) and (\ref{agreennd}) by the first terms of the corresponding series:
\begin{eqnarray}
G(\mathbf{x},\mathbf{x^{\prime}},t)\!\!& \simeq &\!\!\Psi_1(\mathbf{x})\Psi_1(\mathbf{x^{\prime}})e^{-\mu_1^2 D_0t},\\
g(\mathbf{x},T-t)\!\! & \simeq & \!\!\Psi_1(\mathbf{x})\int_{\Omega} d\mathbf{x^{\prime}}\Psi_1(\mathbf{x^{\prime}})e^{-\mu_1^2 D_0(T-t)}.
\end{eqnarray}
This approximation yields the stationary solution
\begin{eqnarray}
\!\!\!q(\mathbf{x})&=& n_0 V\Psi_1^2(\mathbf{x})\label{aproxeqqrwnd}\\
\!\!\!\mathbf{v}(\mathbf{x})&=&\nabla p = \frac{\nabla \Psi_1(\mathbf{x})}{\Psi_1(\mathbf{x})}, \label{aproxeqVrwnd}
\end{eqnarray}
whereas $-\ln {\mathcal P}$ is given by Eq.~(\ref{RWgeneral}).

When $\Omega$ is a circle of radius $R$ ($d=2$), we obtain
\begin{eqnarray}
q (r)&= &\frac{n_0J_0^2\left(\frac{z_1 r}{R}\right)}{J_1^2 (z_1)},\label{q_rw_st2d}  \\
\mathbf{v}(r)&=&-\frac{ z_1 J_1\left(\frac{z_1 r}{R}\right)}{R\,J_0\left(\frac{z_1 r}{R}\right)}\mathbf{\hat{r}},\label{v_rw_st2d}
\end{eqnarray}
where $J_0(z)$ and $J_1(z)$ are Bessel functions, and $z_1=2.4048\dots$ is the first positive root of $J_0(z)$. The survival probability
is described by Eqs.~(\ref{RWresult}) and (\ref{survivaldecay2}).

When $\Omega$ is a sphere  of radius $R$ ($d=3$), the stationary solution is
\begin{eqnarray}
q (r)&= &\frac{2n_0R^2\sin ^2\left(\frac{\pi r}{R}\right)}{3r^2},\label{q_rw_st3d}  \\
\mathbf{v}(r)&=& \left[\frac{\pi}{R}\cot\left(\frac{\pi r}{R}\right) -\frac{1}{r} \right] \mathbf{\hat{r}},\label{v_rw_st3d}
\end{eqnarray}
and the survival probability is described by Eqs.~(\ref{RWresult}) and (\ref{survivaldecay3}).


\begin{thebibliography} {99}
\bibitem{dimensions} In most of our formulas for the SSEP we follow the tradition and put the lattice constant $a$
to unity. The only exception appears at the end of Sec. \ref{SSEP1d}, where we analyse the applicability condition
of the MFT at densities near close packing.
\end{thebibliography}
\end{document}